\documentclass[12pt]{article}
\usepackage{geometry}
\usepackage{graphicx} 
\usepackage{float}
\usepackage{url}
\usepackage{amsmath}
\usepackage{authblk}

\title{Atomistic Simulations of Short-range Ordering with Light Interstitials in Inconel Superalloys}
\author[1,2]{Tyler D. Dole\v{z}al\footnote{corresponding authors: tyler.dolezal.1@us.af.mil; rodrigof@mit.edu; liju@mit.edu}}
\author[3]{Emre Tekoglu}
\author[4]{Jong-Soo Bae}
\author[4]{Gi-Dong Sim}
\author[1]{Rodrigo Freitas$^*$}
\author[1,3]{Ju Li$^*$}
\affil[1]{Department of Materials Science and Engineering, Massachusetts Institute of Technology, Cambridge, MA, USA}
\affil[2]{Department of Engineering Physics, Air Force Institute of Technology, Wright-Patterson Air Force Base, OH, USA}
\affil[3]{Department of Nuclear Science and Engineering, Massachusetts Institute of Technology, Cambridge, MA, USA}
\affil[4]{Department of Mechanical Engineering, Korea Advanced Institute of Science and Technology, 291 Daehak-ro, Daejeon, Yuseong-gu 
34141, Republic of Korea}

\begin{document}

\maketitle

\begin{abstract}
This study employed hybrid Monte Carlo Molecular Dynamics simulations to investigate the short-range ordering behavior of Ni-based superalloys doped with boron or carbon. The simulations revealed that both boron and carbon dissociated from their host Ti atoms to achieve energetically favored ordering with Cr, Mo, and Nb. Boron clusters formed as B\textsubscript{2}, surrounded by Mo, Nb, and Cr, while carbon preferentially clustered with Cr to form a Cr\textsubscript{23}C\textsubscript{6} local motif and with Nb to form Nb\textsubscript{2}C. Distinct preferences for interstitial sites were observed, with boron favoring tetrahedral sites and carbon occupying octahedral sites. In the presence of a vacancy, B\textsubscript{2} shifted from the tetrahedral site to the vacancy, where it remained coordinated with Mo, Nb, and Cr. Similarly, carbon utilized vacancies to form Nb\textsubscript{2}C clusters. Excess energy calculations showed that B and C exhibited strong thermodynamic stability within their short-range ordered configurations. However, under Ti-rich conditions, C was more likely to segregate into TiC, despite preexisting ordering with Cr. This shift in stability suggests that increased Ti availability would alter carbide formation pathways, drawing C away from Cr-rich networks and promoting the development of TiC. Such redistribution may disrupt the continuity of Cr-based carbide networks, which play a critical role in stabilizing grain boundaries and impeding crack propagation. These effects further underscore the impact of interstitial-induced ordering on phase stability and microstructural evolution. This work provides an atomistic perspective on how boron- and carbon-induced ordering influences microstructure and mechanical properties. These findings highlight the critical role of interstitial-induced short-range ordering and demonstrate that this mechanism can be leveraged as a design principle to fine-tune alloy microstructures for specific engineering applications.

\end{abstract}

\section{Introduction}
The materials science community has recently focused on improving the high-temperature mechanical properties of Ni-based superalloys, such as Inconels, by incorporating ultra-high temperature ceramics (UHTCs) as inoculants in solidification and sources of interstitial dopants (B,C) in the metal matrix \cite{fahrenholtzUltrahighTemperatureCeramics2017,niAdvancesUltrahighTemperature2022,wyattUltrahighTemperatureCeramics2023}. While experimental studies have demonstrated the success of light-interstitial dopants in enhancing alloy performance, there remains a gap in understanding their atomistic interactions with transition-metal (TM) solutes (Ni, Cr, Mo, Fe, Nb, Ti, etc.) in the 
superalloy matrix. Specifically, the structural and chemical short-range order (SRO) between light-interstitials and TMs, as well as the impact of light-interstitials on TM-TM SRO, are not fully understood. Computational approaches are essential in bridging this gap, offering detailed insights into the dynamics of interstitials in these systems. These simulations not only reduce the time and cost associated with experimental methods but also enable an in-depth exploration of mechanisms that are challenging to capture experimentally.
\par 
The computational materials science community has increasingly focused on high-temperature applications, using atomistic simulations to explore the fundamental interactions at play \cite{liuApplicationHighthroughputFirstprinciples2021}. Farhadizadeh and Ghomi systematically investigated the mechanical and electronic properties of Zr\textsubscript{x}Ta\textsubscript{8-x}C\textsubscript{8} UHTC, revealing a hardness increase from 22 GPa for TaC to 30 GPa with Zr substitution, and confirmed the presence of mixed covalent, metallic, and ionic bonding via partial density of states and Mulliken charge analysis \cite{farhadizadehMechanicalStructuralThermodynamic2020}. Meanwhile, Wang et al. used first-principles calculations to examine Fe$_2$B, showing that elements like Cr, Mn, and Ni can form continuous solid solutions in (Fe,M)$_2$B, improving its toughness. These findings were validated by experimental results, with Cr- and Ni-doped Fe$_2$B achieving fracture toughness of 4.16 and 4.27\,MPa\,m$^{1/2}$, respectively, compared to 3.64\,MPa\,m$^{1/2}$ for pure Fe$_2$B \cite{wangDesignFe2BbasedDuctile2022}. Additionally, Zhang et al. applied first-principles calculations to explore the structural, electronic, mechanical, and thermodynamic properties of \(\alpha\)- and \(\beta\)-YAlB$_4$, revealing anisotropic elastic moduli, mixed bonding characteristics, and dynamic stability through phonon dispersion analysis \cite{zhangStructuralElectronicMechanical2021}. Finally, Liu et al. explored the impact oxygen interstitials played in CrCoNiFe, predicting site preference and changes in chemical short-range order and mechanical response \cite{liuEffectsShortrangeChemical2020}.  
\par 
Complementing these computational efforts, recent experimental studies have provided valuable insights into the macroscopic behavior of UHTC-dispersed and interstitial-doped superalloys, providing a strong foundation for computational work to support. Yang et al. investigated the effects of intermediate temperature embrittlement (ITE) and the Portevin-Le Châtelier (PLC) effect in Inconel 625 (In625), focusing on how carbide precipitation influenced hot ductility across various temperatures and microstructural evolution during thermal cycling \cite{yangIntermediateTemperatureEmbrittlement2024}. Furthermore, Hu et al. enhanced In718 composites by using ultrasonic vibrations in a graphite mold, improving carbide dispersion and formation, which resulted in superior wear resistance and hardness \cite{huUltrasonicFabricationProcess2024}. Similarly, Teko\u{g}lu et al. demonstrated significant improvements in hardness, strength, and ductility in In718 reinforced with 2\,vol\% ZrB\textsubscript{2} nano-powders using Laser Powder Bed Fusion (LPBF), attributing these enhancements to the formation of homogeneously distributed intermetallic and boride nanoparticles within the matrix \cite{tekogluMetalMatrixComposite2024a}. These experimental advances underscore the potential of UHTC dispersion and interstitial doping for extreme environments.
\par 
Despite recent advances in understanding such fundamental interactions and their impact on superalloy systems, a clear understanding of how interstitials interact with alloy systems, such as Inconels, remains lacking. The mechanisms by which these additives reinforce the alloy matrix, particularly at the atomistic level, are still not fully understood due to complex phenomena such as chemical decomposition and short-range order. Here, we continue to bridge this gap by using atomistic simulations to investigate the microstructural behavior of a mixed-metal simulation cell, mirroring the In625 blend, doped with interstitials -- specifically through the addition of titanium diboride (TiB\textsubscript{2}) and titanium carbide (TiC). An In625 compositional base was selected for this study due to its exceptional mechanical properties, including high strength, corrosion resistance, and oxidation resistance at elevated temperatures \cite{desousamalafaiaIsothermalOxidationInconel2020,chenEffectHeatTreatment2020,chenImprovementHighTemperature2020}. Additionally, In625 has been characterized experimentally, both in its undoped form \cite{desousamalafaiaIsothermalOxidationInconel2020,chenEffectHeatTreatment2020,chenImprovementHighTemperature2020,pariziaEffectHeatTreatment2020,kimHightemperatureTensileHigh2020,sunHightemperatureOxidationBehavior2020,tripathySurfacePropertyStudy2020,ramenatteComparisonHightemperatureOxidation2020,huInfluenceHeatTreatments2021,wangEffectMagneticField2021,sharifitabarHightemperatureOxidationPerformance2022a,guoEffectB4CContent2024,preisEffectLaserPower2024,luoMicrostructureHightemperatureTribological2024}, and when doped with borides or carbides \cite{placeholder}. Therefore, this composition is an ideal candidate for computational investigation where atomistic insights can be experimentally verified.

The performance of In625, particularly when produced through techniques such as Selective Laser Melting (SLM), is highly influenced by its thermal history and heat treatment, which can drastically alter its microstructure and mechanical properties \cite{ferraresi2021,luoMicrostructureHightemperatureTribological2024}. Furthermore, In625 exhibits dynamic recrystallization behavior at high temperatures and low strain rates, which helps maintain its structural integrity under stress \cite{song2021}. Research on cold-rolled In625 foils has further underscored its potential for high-performance applications, demonstrating superior tensile strength and elongation after appropriate annealing \cite{wang2023}. The incorporation of borides/carbides such as TiB\textsubscript{2} and TiC into In625 has shown significant improvements in mechanical performance at elevated temperatures, addressing the limitations of current superalloys and potentially expanding their operational envelope for extreme conditions. While experimental studies have highlighted the potential of borides and carbides in enhancing In625’s high-temperature properties \cite{placeholder}, this work demonstrates, through computational insights, that the light interstitials--specifically boron and carbon (but generalized to include nitrogen and others)--reinforce the alloy via distinct mechanisms. Boron interstitials within the FCC metal matrix induce the formation of small, agile clusters that primarily strengthen grain boundaries, while carbon interstitials provide reinforcement through bulk phase formation. By combining computational results with experimental data, this study offers a first-look at how these light interstitials interact within the host alloy, providing critical insights for optimizing superalloy compositions for energy and aerospace applications.

\section{Methods}
\subsection{Preparing the Bulk Structure}
\begin{figure}[H]
    \centering
    \includegraphics[width=\textwidth]{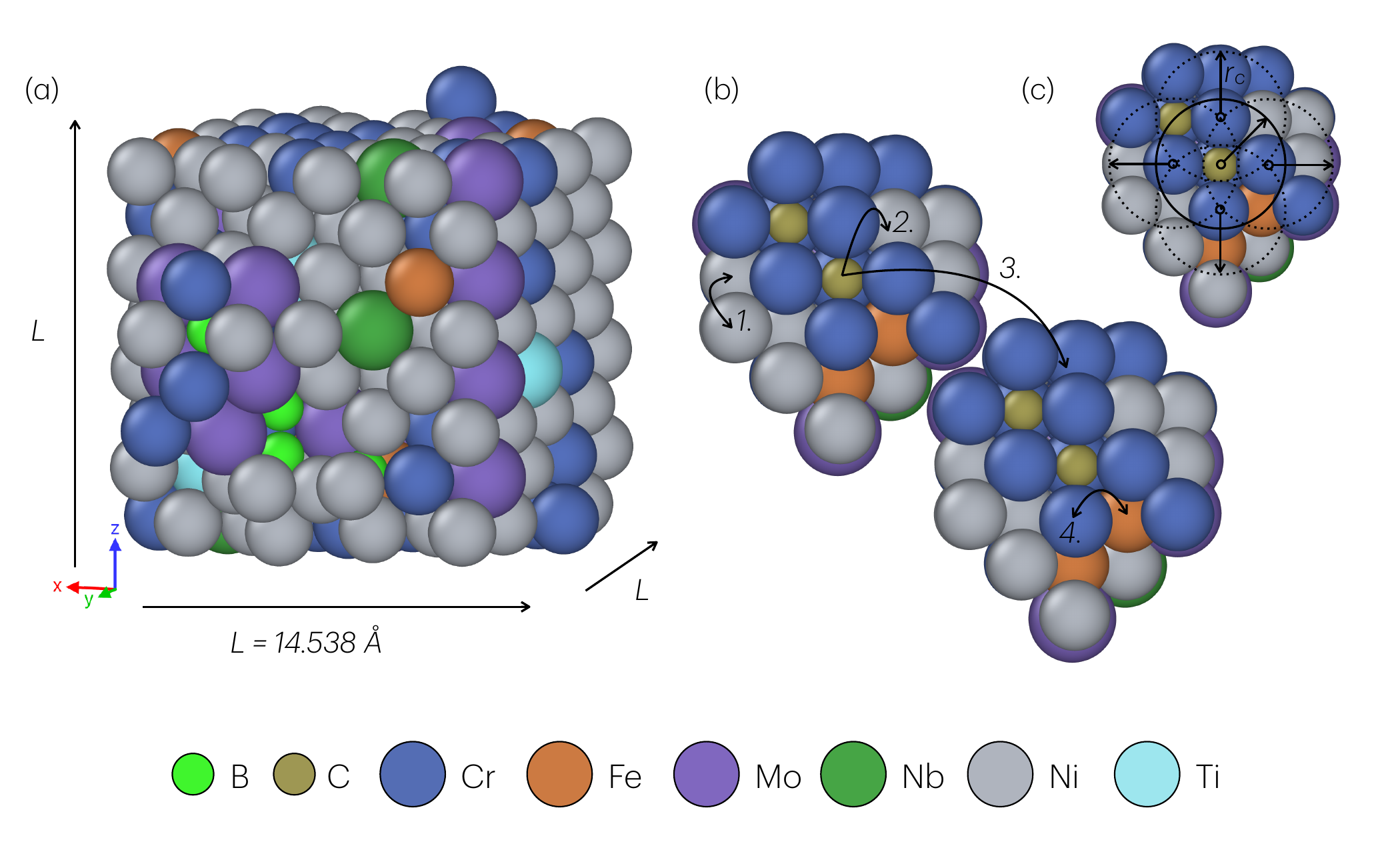}
    \caption{(a) The mixed-metal 4$\times$4$\times$4 FCC lattice which was initially doped with 3 TiB\textsubscript{2} molecules. (b) A schematic as to what each move would consist of. Here, consider each atomic island its own C cluster and the third move would be an attempted addition of a C atom from the top island to the local neighborhood of the C atom on the bottom island. (c) An area within the finalized +TiC mixed-metal system displaying the approximate cutoff distances for local swap or dopant placement events.}
    \label{fig:figure1}
\end{figure}

The bulk structure was initialized as an undoped mixed-metal system generated using the Super-Cell Random Approximation (SCRAP) method in the \(Fm\bar{3}m\) cubic crystal system, consisting of 4$\times$4$\times$4 conventional FCC unit cells, totaling 256 lattice positions. The simulation cell was randomly populated according to the atomic concentration, Cr\textsubscript{22}Fe\textsubscript{5}Mo\textsubscript{10}Nb\textsubscript{3}Ni\textsubscript{58}Ti\textsubscript{2}. For doped systems, 1\,at\% was replaced by the additives (e.g., TiB\textsubscript{2} or TiC) through randomly inserting the additive molecule into an FCC lattice site. This meant a metallic element was randomly selected and switched by the additive molecule (e.g., a Ni atom is switched to a TiB\textsubscript{2}). It was determined that 2 B atoms would occupy a lattice vacancy site once a vacancy was introduced into the simulation cell. The vacancy was introduced by deleting 1 unit of the majority species (Ni in this case). The B-doped simulation cell contained 3 units of TiB\textsubscript{2} (comprising 3 additional Ti atoms and 6 B atoms), while the C-doped cell contained 3 units of TiC (comprising 3 additional Ti atoms and 3 C atoms), along with 3 additional unassociated C atoms to ensure a matched light interstitial concentration with the B-doped system. This resulted in a total of 6 B atoms in the +TiB\textsubscript{2} system and 6 C atoms in the +TiC system. For reference, the final +TiB\textsubscript{2} simulation cell, as seen in OVITO-Open Visualization Tool \cite{stukowskiVisualizationAnalysisAtomistic2009a}, is provided in Fig. \ref{fig:figure1}a. Note that Cr(BMo)\textsubscript{2} molecules formed in what was a Ni vacancy. For this work, there was a single Ni vacancy in the undoped and doped simulation cells, ensuring a consistent structural comparison across the systems.

\subsection{Computational Methods}
Computational investigations were carried out using hybrid Monte-Carlo-molecular-dynamics (hMCMD) simulations. The MC framework of the hMCMD routine would randomly select from a list of four possible local atomic swap or placement events itemized below according to the schematic shown in Fig. \ref{fig:figure1}b.
\begin{enumerate}
     \item \textbf{Swap atomic positions of nearest neighbor metallic constituents:} Select a pair of metallic atoms from within the nearest neighbor shell and attempt to swap their positions in the lattice.
    \item \textbf{Relocate an additive near a new metallic host:} Select an additive atom (B or C) and attempt to place it near a different metallic atom (e.g., randomly) within its nearest neighbor shell.
    \item \textbf{Introduce proximity between two additive atoms (B or C):} Select two additive atoms (B or C) and attempt to place one of them within the first nearest neighbor shell of the other. If the first shell is fully occupied, attempt the placement in the second nearest neighbor shell.
    \item \textbf{Swap a metal neighbor of an additive:} For a selected additive atom (B or C), identify one of its nearest metallic neighbors and swap that metal’s position with one of the metal's own nearest metallic neighbors.
\end{enumerate}
For these moves, nearest neighbors were located within a cutoff radius of 2.75 \AA, which is approximately the regions shown in Fig. \ref{fig:figure1}c. Regardless of which move was chosen, the move was accepted or rejected according to the potential energy difference between the final and initial states following the Metropolis criterion \cite{metropolisEquationStateCalculations1953}. The probability of acceptance is expressed as a Boltzmann probability,
\begin{equation}\label{eq:metropolis}
    P(\Delta U) = \textrm{min}\{1, e^{-\Delta U/{k_\mathrm{B} T_\mathrm{sim}}}\}\textrm{,}
\end{equation}
where $\Delta U$ is the potential energy difference between the final and initial states, $k_\mathrm{B}$ is the Boltzmann constant in eV/K, and $T_{\rm sim}$ is the simulation temperature in K. To align with recent high-temperature experimental efforts \cite{tekogluMetalMatrixComposite2024a, placeholder}, simulations were executed with \(T_{\rm sim}\) set equal to 1073 K (800 \textsuperscript{o}C). In the case of executing the hMCMD simulation on the undoped simulation cell, the first move was the only move trialed. 
\par 
The potential energy was obtained through fixed-volume structural relaxations and MD simulations executed using the Large-scale Atomic/Molecular Massively Parallel Simulator (LAMMPS) software \cite{thompsonLAMMPSFlexibleSimulation2022a}. The calculations utilized version 5.0.0 of the core neural network Preferred Potential (PFP) \cite{takamotoUniversalNeuralNetwork2022} on Matlantis \cite{Matlantis}. Structural relaxations were performed using the conjugate gradient method with energy and force convergence criteria of \(1 \times 10^{-12}\). At the beginning of each simulation, the initial cell was heated to the target temperature of 1073 K (800 \textsuperscript{o}C) and equilibrated for 20 ps under the isothermal-isobaric ($NPT$) ensemble at a pressure of 1 bar, with a time step (\(\delta t\)) of 0.001 ps. Temperature and pressure control during the equilibration process were achieved using a Nose-Hoover thermostat and barostat. The thermostat had a coupling constant of \(100 \times \delta t\), and the barostat had a coupling constant of \(1000 \times \delta t\). After equilibration, a fixed-volume structural relaxation was conducted to finalize the pre-conditioning of the initial simulation cell. At intervals of every 200 successful MC attempts, the simulation cell was subject to a Canonical ($NVT$) MD simulation at $T_\mathrm{sim}$ for a duration of 10 ps with \(\delta t = 0.001\) ps. The temperature was controlled using a Nose-Hoover thermostat with a coupling constant of \(100 \times \delta t\). Following this, 100 MC moves were trialed where potential energy was calculated using short $NVT$ MD simulations of 100 steps with $\delta t$ = 0.001 ps. After 100 MC attempts at the elevated temperature the final structure was subject to a fixed-volume structural relaxation. The final structure was accepted or rejected according to Eq. \ref{eq:metropolis} where the initial state was the configuration that was initially heated to $T_\mathrm{sim}$. Structures from the hMCMD simulation were saved on an interval of every 100 MC attempts. 
\par 
This combined approach of fixed-volume structural relaxations and $NVT$ MD simulations, integrated within the hMCMD routine, enables the exploration of the potential energy landscape of the alloys being investigated. The structural relaxations ensure that the potential energy for MC moves is evaluated at the local energy minima. Meanwhile, $NVT$ MD simulations allow the system to explore configurations at high temperatures, overcoming energy barriers and facilitating transitions between local minima on the potential energy surface (PES). The MC routine further accelerates this process by enabling chemical configuration changes that bypass the kinetic barriers of MD simulations. This method enhances the efficiency of configurational space exploration and leads to structural configurations compatible with thermodynamic equilibrium.

\subsubsection{Clustering Dynamics}

The physical origin of the clustering and short-range ordering observed in this study is rooted in the PES described by the interatomic potential employed in the hMCMD simulations. The potential governs atomic interactions, capturing the energetic balance between attraction and repulsion that dictates clustering dynamics. As stated, this work employs v5.0.0 of the PFP universal neural network potential, supported by Matlantis, which has been extensively validated for modeling Ni-based superalloys and their interactions with light interstitials (B, C) \cite{MatlantisUserTestimonials, PFPValidationPublic2021, mineComparisonMatlantisVASP2023, katoBoronCoordinationThreemembered2024, choungRiseMachineLearning2024, hisamaMolecularDynamicsCatalystFree2024, hinumaNeuralNetworkPotential2024}. This potential has also been rigorously benchmarked against DFT calculations, demonstrating accurate predictions of energy, force, and volume convergence for a wide range of inorganic systems, including their equation of state (EOS) behavior \cite{PFPValidationPublic2021}. Additionally, the shear modulus and Young's modulus predicted for the undoped system (G: 78.79 GPa, E: 209.07 GPa) showed good agreement with experimentally reported values for In625 (G: 78.0 GPa, E: 205 GPa) \cite{specialmetals_inconel625_2013, renishaw_inconel625_2024}.

Light interstitials, such as B and C, can strongly perturb the local PES due to their high chemical affinity for alloy constituents (e.g., Cr, Mo, and Nb). This affinity drives the aggregation of M-B and M-C clusters, as observed in these simulations. The introduction of interstitials modifies the local atomic environments, reducing energy barriers and deepening the PES well, thereby facilitating new SRO patterns. As an example from this work, B preferentially interacts with Cr and Mo, facilitating the formation of Cr(BMo)\textsubscript{2} clusters, while C drives the development of Cr\textsubscript{23}C\textsubscript{6} and Nb\textsubscript{2}C clusters. The clustering trends observed here align well with experimental observations \cite{placeholder, wangNewInsightsPartitioning2025, mignanelliGammagammaPrimegammaDouble2017, goldschmidt4CARBIDES1967, reedSuperalloysFundamentalsApplications2006, zhangSynergyPhaseMC2024, tekogluStrengtheningAdditivelyManufactured2023} and CALPHAD predictions \cite{placeholder}. These findings reinforce the accuracy of the interatomic potential in capturing both atomic-scale mechanisms driving SRO and the elastic response of the alloy, validating its suitability for modeling Ni-based superalloys.

\subsection{Analysis Methods}
\subsubsection{Atomic Ordering and Simulation Statistics}
The average atomic coordination around B and C atoms within the mixed-metal matrix was examined through the partial radial distribution function:
\begin{equation}\label{eq:RDF}
g_{a b}(r) = 4\pi \rho r^2 \left\langle \sum_{i=1}^{N_a} \sum_{j=1}^{N_b} \delta(r - r_{ij}) \right\rangle \quad \text{for} \quad a, b = \textrm{B, Cr, Fe \ldots, Ti}
\textrm{ ,}\end{equation}
where \( \rho \) is the number density of the simulation cell (\(N/V\)), \( N_a \) and \( N_b \) represent the number of atoms of type \(a\) and \(b\), \(\delta\) is the Dirac delta function, \( r_{ij} \) is the distance between atom \(i\) of type \(a\) and atom \(j\) of type \(b\), and the angle brackets \(\langle \cdot \rangle\) denote an ensemble average. For a single species \(a\), this function is calculated for all \(b\) types. While all \( g_{ab}(r) \) functions were calculated, only the case where \(a\) was either B or C is shown to facilitate visualization. This approach enabled the determination of which metallic constituents showed a thermodynamic preference to form clusters with B or C.

Additionally, the average Warren-Cowley short-range order (SRO) parameter ($\alpha_{ij}$) \cite{Cowley1950AnAT, cowley1960} was calculated, as defined in Eq. \ref{eq:SRO},
\begin{equation}\label{eq:SRO}
    \alpha_{ij} (r) = 1 - \frac{P_{ij}(r)}{c_{j}} \textrm{ ,}
\end{equation}
where $P_{ij}$ is the probability of finding atom $j$ as a neighbor to atom $i$ at a distance $r$ and $c_{j}$ is the atomic fraction of the species of atom $j$. These parameters were calculated and averaged using equilibrated structures extracted from simulation snapshots taken at intervals of every 2,500 MC steps after the system had reached equilibrium. Studying the atomic surroundings of B and C atoms from equilibrium is crucial for understanding their interaction dynamics and influence on the alloy's properties. To complement atomic coordination considerations, each species' energetic landscape was examined by evaluating the MC acceptance rate per species normalized by its atomic fraction, as given in Eq. \ref{eq:activity}, 
\begin{equation}\label{eq:activity}
    A_{i} = \frac{C_{i}}{Sx_{i}} \textrm{ ,}
\end{equation}
where $C_{i}$ is the number of times species $i$ has been successfully moved, $S$ is the number of MC steps completed, and $x_{i}$ is the atomic fraction of species $i$. Examining this value provides insight on the relative potential energy difference between the initial and final states. If the MC sampling discovers a very low potential energy state, then there would be a decline in the value of $A_{i}$. For a series of states with near equal potential energy there could be frequent hops between them; in this case, the species would have a higher $A_{i}$ value. Additionally, the evolution of $A_{i}$ over the course of the simulation can be used to show thermal equilibration. The acceptance rate per species ($A_i$) was recorded at an interval of 100 MC steps. 
\subsubsection{Excess Energy}
The excess energy (\(E_{\text{excess}}\)) was calculated as a function of B or C using Eq. \ref{eq:excess}:
\begin{equation}\label{eq:excess}
    E_{\text{excess}}^{(i)} = \frac{E^{(i)}_{\text{doped}} - N_{m}\left(\frac{E_0}{N_0}\right) - N_{i}E_{i}}{N_m + N_{i}} \textrm{ ,}
\end{equation}
where \(E^{(i)}_{\text{doped}}\) is the averaged total energy of the simulation cell doped with interstitial species \(i\), \(N_{m}\) is the number of metal atoms in the doped simulation cell, \(N_{i}\) is the number of light interstitials of species \(i\) in the doped simulation cell, \(E_{0}\) is the energy of the undoped simulation cell, and \(N_0\) is the total number of atoms in the undoped cell. The term \(E_{i}\) represents the energy of the interstitial species in its reference state, which serves as a baseline for determining the relative energetic stability of the dopant within the alloy matrix. The excess energy is a measure of how favorable it is for B or C to remain in the MC equilibrated mixed-metal environment versus leaving to form a reference phase (e.g., B-Mo or C-Cr).

The selection of reference states for the interstitial energy calculation is critical to ensuring a physically meaningful comparison. Since B and C were observed to preferentially cluster into B-Mo and C-Cr configurations during the MC simulations, their energetic contributions were extracted from bulk B-Mo (\(Cmcm\)) and C-Cr (\(Fm\bar{3}m\)) instead of from elemental boron (\(\alpha\)-B) and graphite. Additionally, to assess the influence of alternative boride and carbide phases, reference states for B-Ti (\(P6/mmm\)), B-Ni, C-Ti, and C-Ni (\(Fm\bar{3}m\)) were considered. 

To obtain \(E_i\), the relaxed total energies of the reference states were computed, and the interstitial energy per atom was determined from Eq.~\ref{eq:mu}:

\begin{equation}\label{eq:mu}
    E_{i} = \frac{E_{\text{comp}} - N_{\text{m}} \left(\frac{E_{\text{metal}}}{N_{\text{metal}}}\right)}{N_{i}},
\end{equation}
where \(E_{\text{comp}}\) is the total energy of the reference boride or carbide compound, \(E_{\text{metal}}\) is the total energy of the corresponding bulk metal, and \(N_{\text{metal}}\) is the number of metal atoms in the bulk. In Eq.~\ref{eq:mu}, \(N_{m}\) and \(N_{i}\) represent the number of metal and interstitial atoms in the compound, respectively. The extracted interstitial energies are given in Table \ref{tab:energetics}. By using these reference states, the calculated \(E_{\text{excess}}\) values more accurately capture the thermodynamic stability of B and C in their preferred short-range ordered configurations. This approach ensures that excess energy reflects the balance between interstitial dissolution and clustering within the alloy while also enabling comparisons between different boride and carbide formation tendencies.

\subsubsection{Elastic Properties}
To account for variations in atomic coordination, averages were evaluated using equilibrated structures extracted from simulation snapshots taken at intervals of every 2,500 MC steps after the system had reached equilibrium. Elastic constants \cite{niuExtraelectronInducedCovalent2012,senkovGeneralizationIntrinsicDuctiletobrittle2021a, naherAbinitioStudyStructural2021a, ekumaElasToolV3Efficient2024b} were determined by applying independent deformations along each of the six principal directions ($xx, yy, zz, yz, xz, xy$) in order to build the elastic stiffness tensor. For general elastic response, the bulk ($B$), shear ($G$), and Young's ($E$) modulus were calculated. To examine any changes in ductility between the undoped and doped systems, the Cauchy pressure ($C"$) and Pugh ratio ($G/B$) were calculated: increase in the ductility \cite{senkovGeneralizationIntrinsicDuctiletobrittle2021a,eberhartCauchyPressureGeneralized2012} are often correlated with an increase in $C"$ and a decrease in $G/B$. The Kleinman parameter ($\zeta$) \cite{kleinmanDeformationPotentialsSilicon1962a} was calculated to identify changes in bond behavior; $\zeta \in [0,1]$ is such that higher values indicate a preference for bond bending, while lower values indicate a preference for bond stretching. Together, C" and $\zeta$ can help indicate a trend towards more ductile or brittle behavior.
\subsubsection{Phonon Spectrum}
The high-temperature thermal properties of the undoped and doped systems were explored by comparing their phonon dispersion curves. The dynamical matrix of the system was calculated with LAMMPS using a displacement of 0.02 \AA. The Brillouin zone was sampled along the high-symmetry path of \(\Gamma\)-X-W-K-\(\Gamma\)-L-U-W-L-K at a sample resolution of 50 points between each high-symmetry point. This involved computing the Fourier transform of the dynamical matrix at each interpolated q-point, ensuring proper treatment of the phase factors that arise from the periodic boundary conditions and the positions of atoms within the unit cell. The phonon frequencies were then calculated by diagonalizing the Fourier-transformed dynamical matrix at each q-point. This process was executed on only the final structure sampled from the equilibrated hMCMD simulations.
\subsubsection{Bader Charge Analysis}
Bader charge analysis was performed to investigate the electronic charge distribution within the simulated systems. The calculations were conducted using the Vienna Ab initio Simulation Package (VASP) \cite{kresseEfficientIterativeSchemes1996,kresseUltrasoftPseudopotentialsProjector1999} with the Perdew-Burke-Ernzerhof (PBE) exchange-correlation functional \cite{perdewGeneralizedGradientApproximation1996} within the generalized gradient approximation (GGA). The projector augmented wave (PAW) method \cite{kresseUltrasoftPseudopotentialsProjector1999} was employed with a plane-wave energy cutoff of 520 eV. A Monkhorst-Pack \cite{monkhorstSpecialPointsBrillouinzone1976} k-point grid of 4$\times$4$\times$4 was used to sample the Brillouin zone. The self-consistent field calculations were continued until the total energy difference between successive iterations reached a threshold of 
\(1\times10^{-6}\) eV. The calculations employed the ``high quality'' B and C pseudopotentials, which had valence electron configurations of 2s\textsuperscript{2}2p\textsuperscript{1} and 2s\textsuperscript{2}2p\textsuperscript{2}, respectively. The metallic elements had valence configurations as follows: Cr with 3d\textsuperscript{5}4s\textsuperscript{1}, Fe with 3d\textsuperscript{7}4s\textsuperscript{1}, Mo with 4d\textsuperscript{5}5s\textsuperscript{1}, Nb with 4d\textsuperscript{4}5s\textsuperscript{1}, Ni with 3p\textsuperscript{6}3d\textsuperscript{8}4s\textsuperscript{2}, and Ti with 3d\textsuperscript{3}4s\textsuperscript{1}. This work used the pseudovalence (pv) potentials, which include electrons from orbitals typically considered as core (e.g., 3p for Ni). The use of these potentials was particularly appropriate given the need to capture the subtle electronic effects of alloying and doping in such a highly alloyed material system. The Bader analysis was carried out on the final hMCMD simulation cells using the Bader Charge Analysis code \cite{henkelman2006fast, sanville2007improved, tang2009grid, yu2011accurate}.
\section{Results and Discussion}
\subsection{Atomic Ordering}
\begin{figure}[H]
    \centering
    \includegraphics[width=\linewidth]{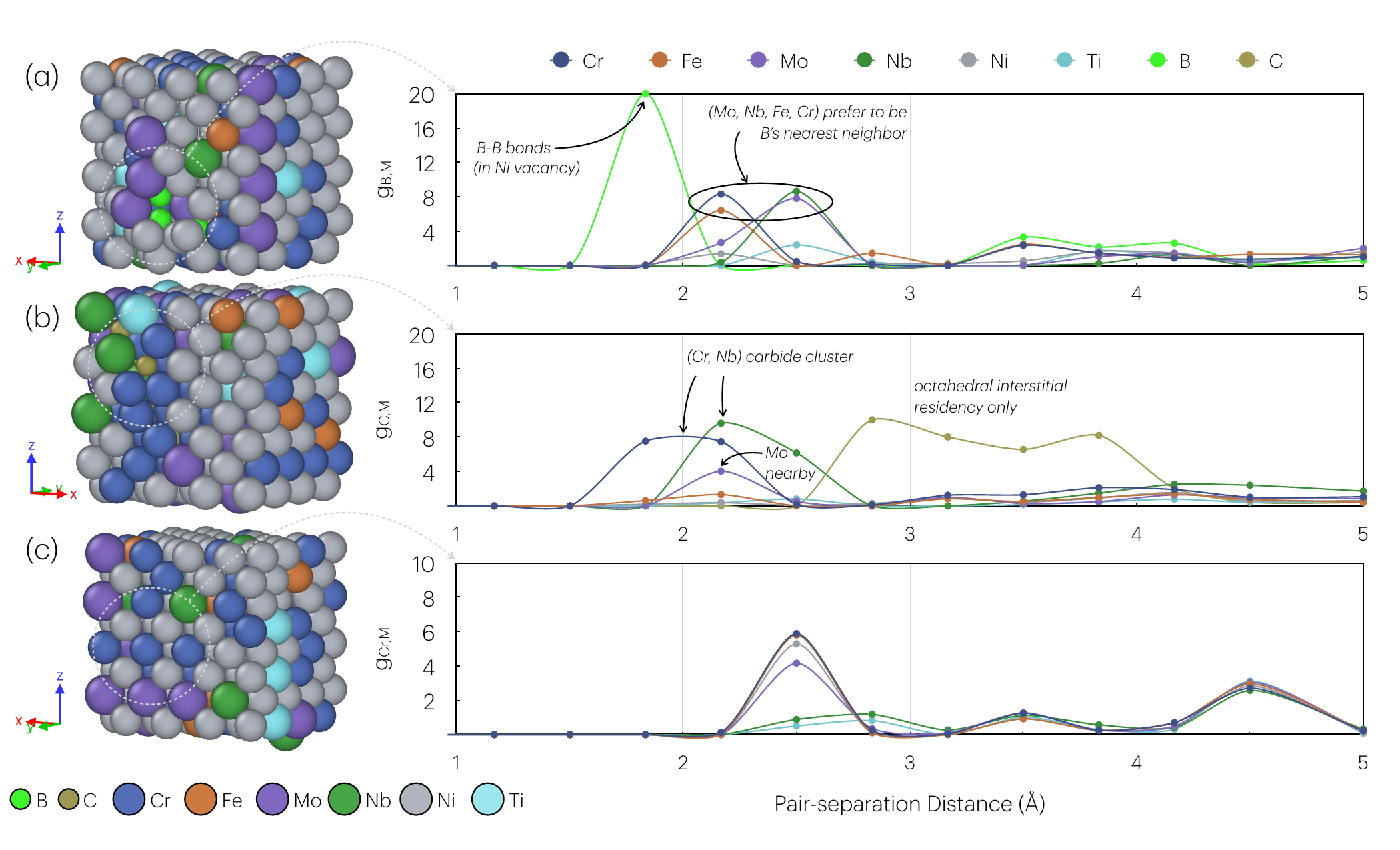}
    \includegraphics[width=\linewidth]{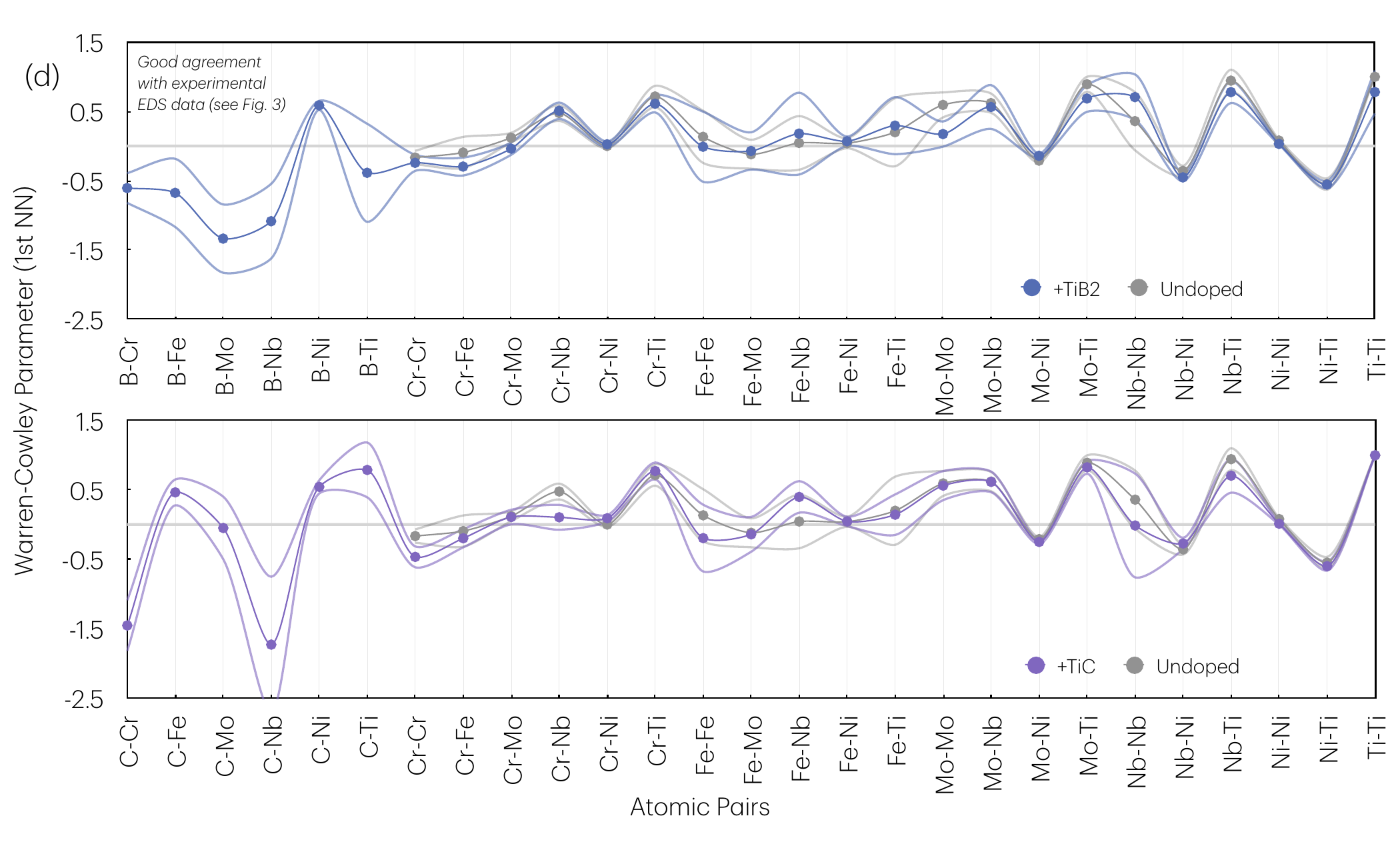}
    \caption{The final (a) +TiB\textsubscript{2}, (b) +TiC, and (c) undoped mixed-metal simulation cells from the hMCMD simulation for $T_{\rm sim}$ equal to 1073 K (800 \textsuperscript{o}C). The B and C clusters have been encircled with approximately the same volume outlined in the radial distribution function curves. (d) Warren-Cowley parameters for the first nearest-neighbor shell. The standard deviation is represented by a veil of uncertainty around the solid data points.}
    \label{fig:ordering}
\end{figure}
\subsubsection{Boron-doped System}
Figure \ref{fig:ordering}a illustrates the final system geometry of +TiB\textsubscript{2}. In all simulations, B\textsubscript{2} clusters formed within Ni vacancy sites and were consistently surrounded by at least two Mo atoms. In the absence of the Ni vacancies, B\textsubscript{2} formed in the tetrahedral interstitial sites. Once these clusters formed, both the B and Mo atoms exhibited negligible mobility, indicating a very low potential energy configuration. The environment surrounding B\textsubscript{2} was rich in Cr, Mo, and Nb, as demonstrated by the RDF graph in Fig. \ref{fig:ordering}a and the SRO analysis in Fig. \ref{fig:ordering}d. Particular attention should be given to the RDFs for the light interstitials (B and C), as they reveal the ordering tendencies and provide insights into the structural characteristics from the perspective of interstitial sites. This distinction becomes evident when compared to the RDF for Cr-M, which represents a more conventional on-lattice RDF, as shown in Fig. \ref{fig:ordering}c. Given that B was introduced into the lattice through the compound TiB\textsubscript{2}, and new ordering is observed in the final structures, it is evident that the B\textsubscript{2} dissociates to achieve a more energetically favorable configuration and order.

\begin{figure}[H]
    \centering
    \includegraphics[width=\linewidth]{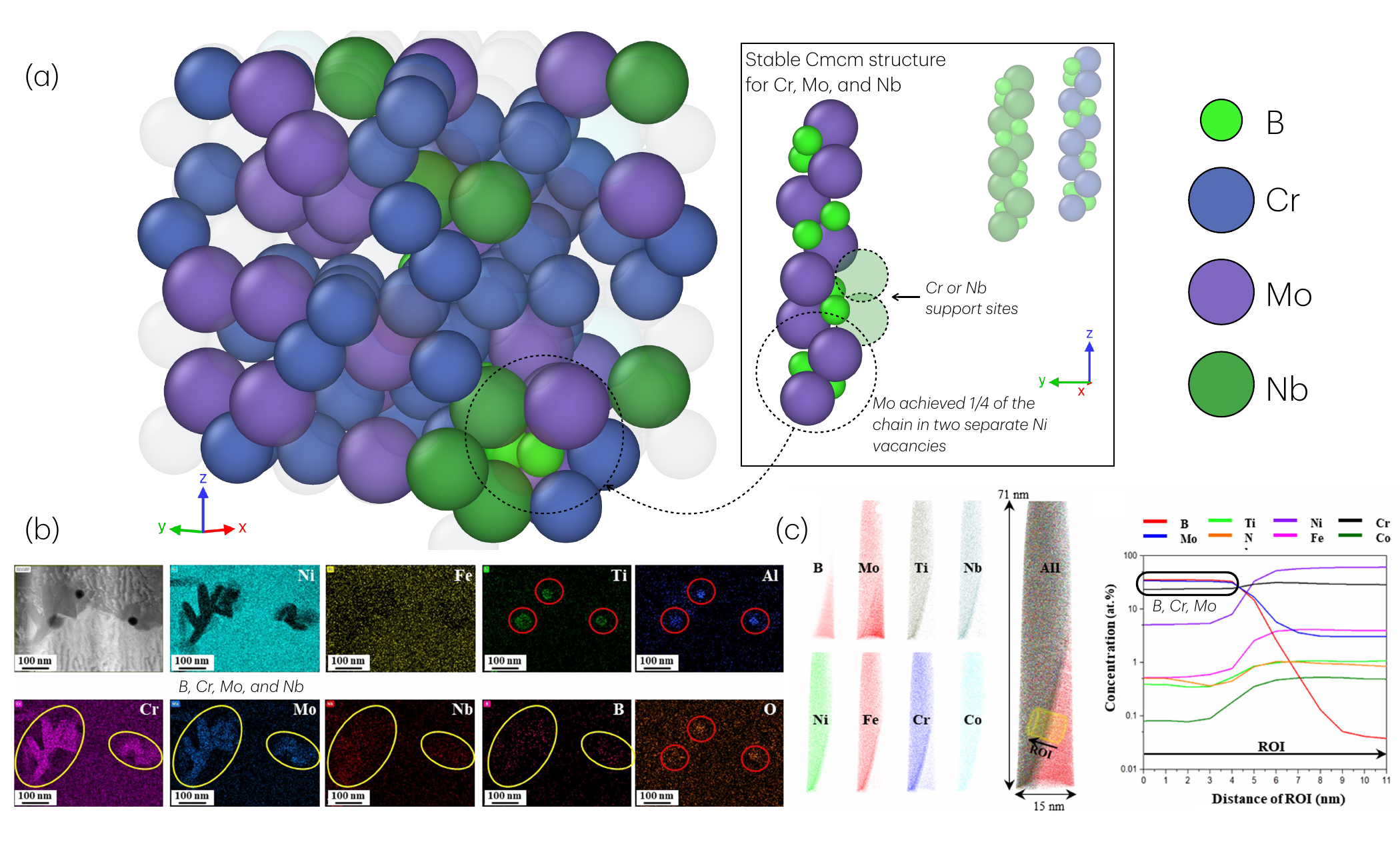}
    \caption{(a) The final +TiB\textsubscript{2} simulation cell with Ni, Fe, and Ti atoms made transparent to facilitate the visualization of one of the B$_2$ clusters. Alongside this are illustrations of the Cmcm (Cr, Mo, or Nb)-B structure. (b) EDS micrographs from an In625+TiB\textsubscript{2} sample, highlighting good ordering between Cr, Mo, Nb, and B. (c) Atom probe tomography data extracted from a deformed sample of In625+TiB\textsubscript{2}, showing preferential ordering between Cr, Mo and B. Subfigure (b-c) was reproduced with permission from Ref. \cite{placeholder}.}
    \label{fig:gbs}
\end{figure}

It is noteworthy that Cr, Mo, and Nb can form the same stable Cmcm structure with B \cite{kayhanTransitionMetalBorides2013, okadaStructuralInvestigationCr2B31987}, as illustrated in Fig. \ref{fig:gbs}a. In this simulation, the B\textsubscript{2} cluster constituted one-quarter of the related Cmcm structure. A comparison between the ordering predictions from this simulation and recent Energy Dispersive Spectroscopy (EDS) micrographs of an In625+TiB\textsubscript{2} sample, shown in Fig. \ref{fig:gbs}b, demonstrates excellent agreement with real-world observations \cite{placeholder}. This consistency suggests that the predicted dissociation of TiB\textsubscript{2} and the resulting energetically favorable ordering will likely occur in alloys with sufficient Cr, Mo, and Nb. Additionally, atom probe tomography (APT) data from a deformed (post-creep test) In625+TiB\textsubscript{2} sample, presented in Fig. \ref{fig:gbs}c, further supports this conclusion. The preferential ordering between Cr, Mo, and B observed in the APT data aligns closely with the predictions of this study, indicating that the SRO reported here remains intact even under mechanical stress.

Recent experimental works provide compelling evidence on the partitioning behavior of Mo and Cr in Ni-based superalloys, showing that Cr preferentially partitions to the \(\gamma\) matrix while Mo’s distribution shifts dynamically with Cr content \cite{wangNewInsightsPartitioning2025, liuEffectMoAddition2015, wangInfluenceReCr2016, maPartitioningBehaviorLattice2021}. Higher Cr concentrations reduce Mo’s solubility in \(\gamma'\) precipitates, driving its redistribution to the \(\gamma\) phase. Building on these observations, this work reveals that B-rich environments further influence Cr and Mo behavior by promoting the formation of Cr(BMo)\textsubscript{2} clusters. Identified here as energetically favorable, such structures are likely to occur in Ni superalloys with sufficient Cr and Mo concentrations, highlighting the competitive interactions between Cr and Mo under these conditions. Furthermore, these studies \cite{wangNewInsightsPartitioning2025, liuEffectMoAddition2015, wangInfluenceReCr2016, maPartitioningBehaviorLattice2021} also highlighted the influence of \(\gamma\)/\(\gamma'\) lattice mismatch on phase stability and morphology, emphasizing how elemental distribution impacts microstructural evolution and mechanical performance. Given that lattice mismatch plays a crucial role in phase stability, the preferential aggregation of Cr and Mo around B could influence the lattice mismatch at grain boundaries, potentially enhancing the stability of adjacent phases. In this context, the aggregation of Cr and Mo around B may suggest that such clusters can serve as nucleation sites for phase stabilization, particularly at grain boundaries. This behavior has significant implications for improving high-temperature performance and creep resistance in B-doped Ni-based superalloys \cite{placeholder, wangInsightLowCycle2023}.

The interplay between Cr, Nb, and B observed in this work reflects a competitive dynamic that significantly influences microstructural evolution and phase stability. Mignanelli et al. \cite{mignanelliGammagammaPrimegammaDouble2017} demonstrated that elemental partitioning in systems containing \(\gamma\), \(\gamma'\), and \(\gamma''\) phases dictates the formation and coexistence of precipitates, with Cr favoring the \(\gamma\) matrix and Nb enriching the \(\gamma''\) phase. The findings of this study suggest that the presence of B further complicates these partitioning behaviors by modifying local atomic environments and influencing mechanical properties through its impact on dislocation dynamics. Wang et al. \cite{wangInsightLowCycle2023} observed that B enhances dislocation entanglement near \(\gamma\)/\(\gamma'\) interfaces, reducing stress localization and improving fatigue resistance. Similarly, Cr’s tendency to drive Nb out of \(\gamma\) into \(\gamma''\) aligns with competitive partitioning behaviors between Cr and Mo reported in previous works \cite{wangNewInsightsPartitioning2025, liuEffectMoAddition2015, wangInfluenceReCr2016, maPartitioningBehaviorLattice2021}. Building on these insights, this work demonstrates that B stabilizes Cr and Nb within \(\gamma\)-rich environments, such as the mixed-metal simulation cell, which may have implications for grain boundary stability and creep resistance. These findings provide a framework for understanding the competitive interactions in the B-doped system, where Cr, Mo, Nb, and B collectively drive short-range ordering and microstructural evolution.

The interactions highlighted in this study offer interesting insights into the behavior and localization of B atoms in B-doped Ni superalloys. As suggested in this work, B’s high chemical affinity for Cr, Mo, and Nb should lead to localized ordering, stabilizing Cr-Mo-Nb-B clusters in regions enriched with these elements. The preference of Nb for the \(\gamma''\) phase and Cr for the \(\gamma\) phase suggests that B may preferentially associate with Nb-enriched \(\gamma''\) regions, further enhancing the stability of Cr-Nb-B clusters. Furthermore, B’s ability to pin dislocations near grain boundaries \cite{wangInsightLowCycle2023} reinforces its role in improving grain boundary stability and resistance to thermally activated creep deformation \cite{placeholder}. These interactions could be further influenced by local compositional variations, mediating phase stability and suppressing deleterious transformations such as the formation of the \(\delta\) phase (Ni\textsubscript{3}Nb). The dual-superlattice structure discussed by Mignanelli et al., with comparable fractions of \(\gamma'\) and \(\gamma''\) phases, underscores the potential for tailored compositions to enhance both thermal stability and mechanical performance. Incorporating B into these systems may further enhance the stability of \(\gamma''\), potentially impeding the coarsening of Nb-rich precipitates and promoting Cr, Mo, and Nb aggregation near grain boundaries or within specific phases. These behaviors collectively play a critical role in improving creep resistance and maintaining high-temperature mechanical integrity in advanced Ni-based superalloys.

\subsubsection{Carbon-doped System}

Switching focus to the M-C clusters, Fig. \ref{fig:ordering}b illustrates the atomic coordination for the +TiC simulation cell, with the SRO of the C-doped system shown in Fig. \ref{fig:ordering}d. Due to the larger size of C atoms, they exclusively occupied octahedral interstitial sites. Similar to the B-doped cell, the introduction of Ni vacancies significantly altered the nearest-neighbor preference. In the +TiC system, a strong interaction between Cr, Nb, and C was observed. Notably, the presence of Ni vacancies facilitated the formation of the Nb\textsubscript{2}C cluster, as shown in the final results. Without these vacancies, Nb was unable to form stable Nb\textsubscript{2}C clusters. As observed with TiB\textsubscript{2}, TiC dissociated to enable the formation of more energetically favorable ordering. The introduction of C to the simulation cell promoted significant clustering of Cr and Nb around C atoms. These findings align well with commonly reported carbide precipitates in a wide range of Ni-based superalloys \cite{yangIntermediateTemperatureEmbrittlement2024, huUltrasonicFabricationProcess2024, goldschmidt4CARBIDES1967, reedSuperalloysFundamentalsApplications2006, zhangSynergyPhaseMC2024}. High concentrations of Cr drove the partial formation of the Cr\textsubscript{23}C\textsubscript{6} structure, which adopts the Fm$\bar{3}$m symmetry \cite{lvStructuralPropertiesPhase2014}, while the Nb\textsubscript{2}C cluster oriented itself in the Pnma symmetry \cite{yvonKristallstrukturSubcarbideUebergangsmetallen1967}. However, due to the fixed atomic concentration and limited space in the 4$\times$4$\times$4 FCC supercell, these clusters were unable to fully segregate and precipitate as they would in a real-world sample.

The deformation behavior of carbide-containing systems highlights the critical influence of carbide morphology on mechanical performance. As noted by Li et al. \cite{liInfluenceCarbidesPores2024}, script-like carbides mitigate strain by dispersing lamellar carbides within the $\gamma$ and $\gamma'$ phases, reducing slip band formation. In contrast, rod-like and flake-like carbides obstruct dislocation motion, causing strain localization that often leads to carbide fracture, debonding, and crack initiation. Similarly, additional studies emphasized the critical role of carbide morphology and distribution in determining mechanical performance \cite{zhangSynergyPhaseMC2024,wangInsightLowCycle2023, fengEffectPreAddedHfO22024}. With the reported observations that fine M\(_{23}\)C\(_6\) carbides pin dislocations in the $\gamma$ channels, stabilizing the matrix and delaying crack propagation, while fragmented carbides act as crack initiation sites due to dislocation accumulation and higher interface energy \cite{suComputationalStructuralModeling2014}. Karamched and Wilkinson further highlighted that stress gradients near carbides amplify geometrically necessary dislocations accumulation, particularly near sharp corners, which could exacerbate stress localization \cite{karamchedHighResolutionElectron2011}. Such behaviors are particularly relevant in the C-doped system, where Cr-C and Nb-C clustering indicate a strong potential for carbide-driven microstructural evolution. These findings align with prior reports, suggesting that Cr and Nb clustering around C atoms could lead to the development of fine carbides in $\gamma$ channels, thereby mitigating crack propagation and improving mechanical performance. This clustering behavior highlights the critical interplay of alloy composition and microstructural stability in deformation environments, underscoring the importance of controlling carbide morphology and distribution for optimized high-temperature performance.

The synergy between carbides, \(\gamma'\) phases, and grain boundaries, as emphasized by \cite{zhangSynergyPhaseMC2024, wangInsightLowCycle2023}, suggests that stable Cr\textsubscript{23}C\textsubscript{6} clusters play a vital role in stabilizing grain boundaries \cite{suComputationalStructuralModeling2014} and enhancing creep resistance. Additionally, Wang et al. \cite{wangInsightLowCycle2023} observed that C reduces stacking fault energy, promoting planar slip and localized deformation. By reducing fragmented carbide density, these clusters contribute to continuous carbide networks, mitigating crack initiation \cite{zhangSynergyPhaseMC2024, wangInsightLowCycle2023}. Additionally, homogeneous carbide distributions improve plasticity and delay crack propagation by redistributing stress throughout the matrix \cite{zhangSynergyPhaseMC2024, wangInsightLowCycle2023}. Given the clustering behavior observed here and recent reports of improved high-temperature performance in C-doped Inconels \cite{yangIntermediateTemperatureEmbrittlement2024, huUltrasonicFabricationProcess2024,placeholder}, C doping appears to be a key driver of carbide evolution and mechanical stability in Ni-based superalloys. Ultimately, carbides influence mechanical properties based on their morphology and distribution. Well-distributed carbides reinforce the matrix by hindering dislocation movement, whereas fragmented carbides can localize stress and accelerate failure. These findings underscore the importance of controlled C doping in optimizing carbide stability and high-temperature performance in Ni-based superalloys.

\subsubsection{The Interplay Between Boron- and Carbon-Induced Ordering}
Both M-B and M-C structures with 4d transition metals (Mo and Nb) required the flexibility provided by vacancies to achieve their preferred geometries, while the Cr-rich C structure remained intrinsically stable within the FCC matrix. A key distinction between the M-B and M-C clusters lies in their behavior within the host alloy. For instance, the M-B\textsubscript{2} clusters observed here preferentially form along grain boundaries, stabilizing their structure and influencing grain boundary morphology \cite{placeholder,wangInsightLowCycle2023}, whereas Cr-C and Nb-C clusters aggregate to form complex networks of various morphologies \cite{zhangSynergyPhaseMC2024, wangInsightLowCycle2023}. Considering that carbides are common precipitates in Ni-based superalloys, the interaction between B and pre-existing carbides warrants further discussion.

This study demonstrates that B preferentially interacts with carbide-forming elements such as Cr, Mo, and Nb, suggesting the potential formation of mixed boro-carbides, such as M\textsubscript{23}(B, C)\textsubscript{6} \cite{singhRolesRefractorySolutes2024}. These interactions can refine carbide morphology and stabilize grain boundaries, as observed in recent works \cite{singhRolesRefractorySolutes2024, theskaReviewMicrostructureMechanical2023, kangMicrostructuralAnalysisGrain2024}. Notably, B influences carbide precipitation kinetics, often resulting in refined and more stable carbide structures that enhance mechanical properties. However, Singh et al. \cite{singhRolesRefractorySolutes2024} revealed that alloying elements such as Ti and W can destabilize mixed boro-carbides, thereby limiting their positive contributions to grain boundary stability and overall mechanical performance. The competitive interactions between B and C at grain boundaries are particularly significant, as they influence the distribution and morphology of carbides. Improper doping or destabilizing alloying elements may lead to fragmented or poorly distributed carbides, exacerbating stress localization and reducing their reinforcing effects \cite{zhangSynergyPhaseMC2024, wangInsightLowCycle2023, liInfluenceCarbidesPores2024, fengEffectPreAddedHfO22024}. Understanding the interplay between M-B and M-C clusters, especially in the context of interactions with other alloying elements, is critical for the design of advanced Ni-based superalloys. The atomistic methods employed in this study provide a powerful tool to identify elements that may stabilize or destabilize beneficial M-B and M-C clusters. By leveraging these insights, alloy compositions can be optimized to promote the formation of stable borides, carbides, or mixed boro-carbides, aimed at enhancing the mechanical stability and high-temperature performance of Ni-based superalloys.

\subsubsection{Short-range Order Relative to the Undoped System}
The undoped system exhibited both aggregation and segregation behaviors among its metallic constituents. Interestingly, Cr-Cr showed a preference to aggregate, despite Cr's known tendency to avoid itself in a Ni lattice \cite{sheriffQuantifyingChemicalShortrange2024}. This highlights how clustering dynamics shift with increased chemical complexity. Introducing B or C into the mixed-metal simulation cell generated new SRO among metal-metal (M-M') pairs, further altering clustering dynamics between metallic constituents in the presence of light interstitial dopants.

In the B-doped system, aggregation involving Cr, Fe, Mo, and Nb was observed around the solute atoms. Due to the low concentrations of Mo and Nb in the cell, capturing this change accurately in the $\alpha_{MM'}$ values is challenging. However, undoped SRO values suggest that without B, these species tend to avoid one another. With only 1 at\% B present, the promotion of clustering remains limited but significant. In the C-doped system, the partial formation of the M\textsubscript{23}C\textsubscript{6} M-C structure promoted Cr-Cr aggregation (indicated by the more negative $\alpha_{CrCr}$ value), while the formation of Nb\textsubscript{2}C enhanced Nb-Nb clustering, despite Nb's usual tendency to avoid itself (as indicated by the $\alpha_{NbNb}$ parameter). Notably, there was clear avoidance between both B and C with Ni, suggesting that it should be uncommon to find these interstitials in the \(\gamma\) matrix.

Relative to the undoped system's SRO, and accounting for conservation of atomic concentrations, the addition of B or C induced new SRO patterns around the solute atoms. This, in turn, altered the metallic clustering dynamics in their local environments, particularly in interactions with 4d transition metals, which have been shown to exhibit their own influence on one another \cite{wangNewInsightsPartitioning2025, liuEffectMoAddition2015, wangInfluenceReCr2016, maPartitioningBehaviorLattice2021}. These new M-M' clustering dynamics are expected to evolve further with higher B or C concentrations. Wei et al. \cite{weiOneStepOxides2024} corroborate these findings, showing that interstitials such as B and C can induce new SRO in metal matrices through elemental redistribution, leading to new microstructural arrangements and phase stability. Furthermore, in addition to the experimental works cited, the observed SRO aligns well with CALPHAD predictions \cite{placeholder}, reinforcing the validity of this method.

\subsection{Simulation Statistics}
\begin{figure}[H]
    \centering
    \includegraphics[width=\linewidth]{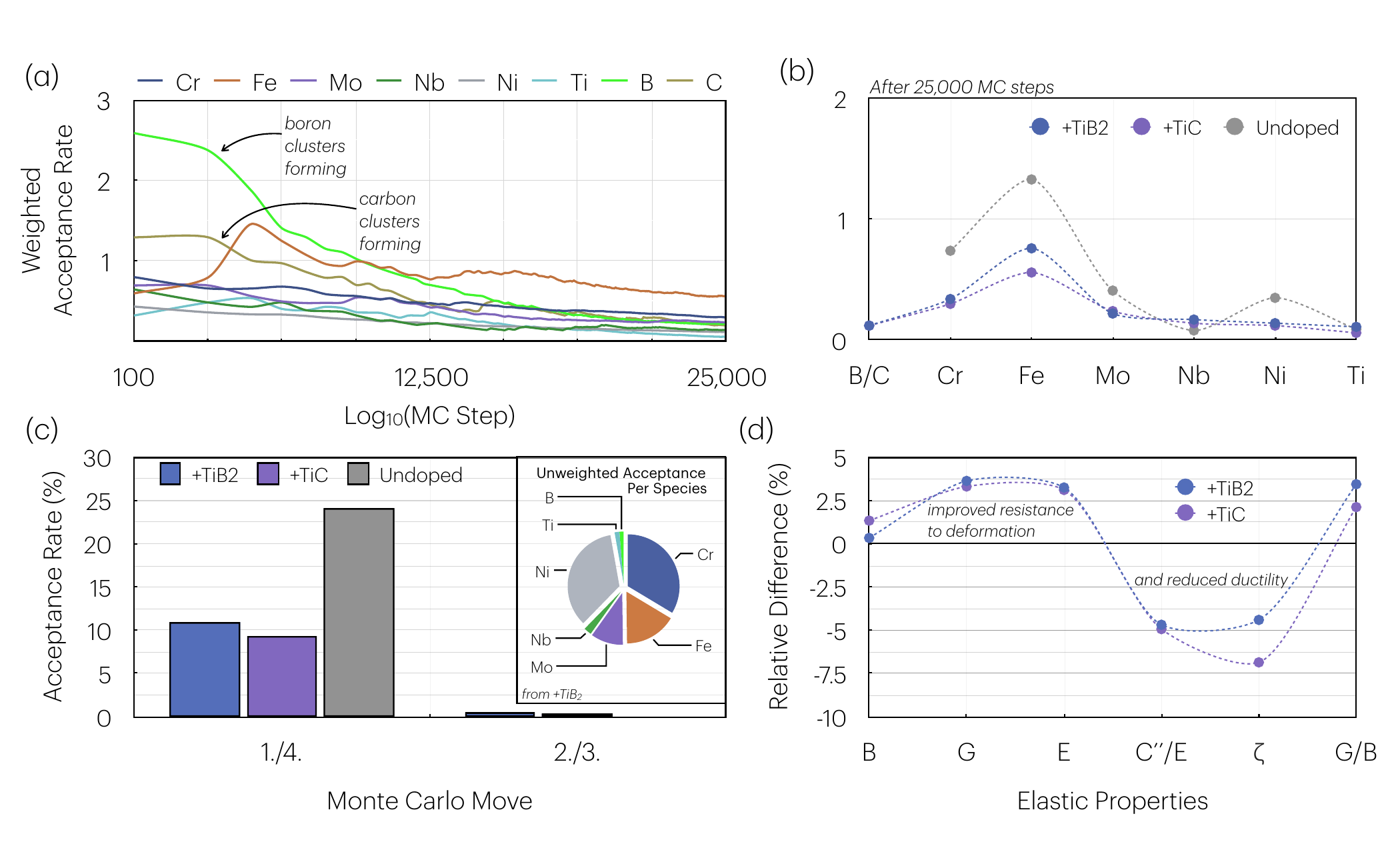}
    \caption{(a) The weighted acceptance rate of each species plotted as a function of MC step on a $\log_{10}$ scale, with x-axis labels corresponding to the actual MC step. By examining the weighted acceptance over the lifetime of the simulation it is possible to see a cluster-formation epoch followed by a steady-state epoch where atomic swaps slow in response to the creation of the M-B or M-C clusters. Metallic curves were generated from the +TiC simulation data. (b) The total weighted acceptance rate of each constituent for the different systems after 25,000 MC steps. A higher value indicates a species was more mobile in the lattice. (c) The total acceptance rate for the undoped and doped systems split between MC move types where moves 1 and 4 involved metal swaps and moves 2 and 3 involved B or C movement. The inset figure was taken from the +TiB\textsubscript{2} simulation statistics and displays the unweighted (e.g., not divided by atomic fractions) look at which species contributed most to the overall atomic motion. (d) The averaged elastic properties for the B-doped system (+TiB\textsubscript{2}) compared against the C-doped system (+TiC) expressed as their properties relative to the undoped system's average elastic properties at $T = 800$ \(^{\circ}\)C. Variation in the elastic properties was minimal (about 1\%) over the different SRO configurations.}
    \label{fig:dynamics}
\end{figure}
Figure \ref{fig:dynamics} presents a collection of graphs that displays the statistics from the hMCMD simulation for the Ni-superalloy-based mixed metal system with and without the B or C atoms. It is important to recall that the MC moves involved the movement of either a metallic constituent or a dopant atom throughout the lattice via swapping or hopping in their nearest neighbor shell. By keeping track of how many times an atom of each type successfully swapped or hopped, one can investigate the PES from the perspective of each species. 
\subsubsection{Weighted Acceptance Rate over Sim Time}
Figure \ref{fig:dynamics}a shows each species' activity level over the entire hMCMD simulation for the +TiC system with the inclusion of the B curve from the +TiB\textsubscript{2} simulation. It reveals that cluster formation occurred very early in the simulation, when B and C were most active, followed by a sharp drop in their mobility as the M-B and M-C clusters were established. After the clustering phase, the system transitioned into a steady-state phase characterized by metallic hopping. This is a good indication of a thoroughly equilibrated simulation. Interestingly, Fe remained particularly active throughout the simulation. In fact, Fe showed the highest activity in all the systems (see Fig. \ref{fig:dynamics}b). This suggests that Fe should be found more uniformly throughout the matrix and an experimental examination of the In625+TiB\textsubscript{2} microstructure did show a nearly uniform distribution of Fe in the In625 matrix \cite{placeholder}. For the B-doped system, most of the swaps observed in the later stages of the hMCMD simulations involved the reorganization of metallic constituents in the simulation cell and around the B\textsubscript{2} clusters. Meanwhile, in the C-doped system, atomic swaps or new placement of C atoms within and near the M-C cluster were very limited. This suggests that the PES was more strongly perturbed in the presence of the M-C structure than the M-B structure.
\par
Given that the MC moves were restricted to local movement, the quick formation of the B and C clusters within the lattice may provide insights into the diffusion dynamics and the stabilization of the material’s microstructure. The formation of these clusters early in the simulations suggests a strong thermodynamic driving force for B and C atoms to aggregate and form stable M-B and M-C clusters. The quick clustering followed by a dramatic reduction in movement indicates that the energy landscape around these atoms was steep, leading to an early “lock-in” of the clusters. Interestingly, a recent ultrasound study on In718 and carbide phase development suggested that applying ultrasound during processing can slow down the rapid clustering of C, allowing more time for the formation of varied carbide phases, which were dispersed more uniformly to produce a harder alloy \cite{huUltrasonicFabricationProcess2024}. Nevertheless, the strong thermodynamic drive for the formation of the M-B and M-C clusters could suggest that B or C reinforced Ni superalloys are well-suited to the conditions of additive manufacturing (AM), where rapid solidification and precise microstructural control are critical. Timely clustering behavior may not only stabilize the microstructure during the AM process, potentially reducing the need for extensive post-processing, but also allow for tailoring material properties to meet specific application requirements. This could suggest B and/or C reinforced Ni superalloys are particularly viable for use in AM techniques, however future work incorporating time-dependent diffusion modeling will be necessary to fully explore this aspect.
\subsubsection{Total Weighted Acceptance Rate}
Fig. \ref{fig:dynamics}b shows the final weighted acceptance rate (e.g., after 25,000 MC steps) of each species in the different systems. Notably, the mobility of atoms was reduced in the doped systems. This suggests that the formation of the M-B or M-C clusters established a region within the FCC lattice that reduced atomic mobility. It is important to note that the undoped system showed the highest atomic mobility values. This could point to a common tendency for dynamic reordering among blends of Ni superalloys, which facilitates stress redistribution through atomic movement. This mobility could make the undoped system less resistant to long-term mechanical loads or high temperatures, as there are no diffusion barriers like those created by the M-B or M-C clusters. A slight increase in Nb mobility suggests that the addition of B or C to the lattice formed an energy surface that enhanced Nb's mobility in the lattice.
\par 
The observed difference in metallic activity near the M-B or M-C structure reveals distinct properties. For instance, metallic swaps within or nearby the region occupied by the large Cr-C motif was limited, suggesting that the M-C clusters formed stable structures which resisted further atomic reordering. While this stability can enhance hardness and wear resistance, it could also lead to brittleness and an increased susceptibility to microcrack formation under stress. In contrast, metallic reordering continued in the vicinity of the M-B clusters, suggesting greater flexibility and accommodation of atomic movement. This may indicate M-B clusters can redistribute stresses more effectively, potentially preventing the formation of stress concentration points and enhancing the material's toughness and resistance to crack propagation.
\subsubsection{Insights from the Acceptance Rates}
Mobility was quantified through the acceptance rate which depended on the potential energy difference between the initial and final state. The results suggest that the energetic landscape around the B and C cores significantly influenced metallic swap and new dopant placement dynamics. Once the (BMo)\textsubscript{2} cluster formed, new placement or swaps of those B and Mo atoms ceased. The same was true for Cr and Nb once it occupied nearest-neighboring sites around a C atom. Figure \ref{fig:dynamics}c compares the final acceptance rates for MC moves 1 and 4 (metallic atom moves) with MC moves 2 and 3 (B or C atom moves). A sharp reduction in metallic moves was observed in systems containing B or C atoms, further supporting the notion that the addition of B or C to the interstitial medium slowed atomic reordering. As previously noted, metal atoms showed higher mobility around the M-B clusters compared to M-C clusters, which is reflected in the higher acceptance rate for moves 1 and 4 in the +TiB\textsubscript{2} system versus the +TiC system. Approximately 0.5\% of attempts to move a B atom succeeded, while fewer than 0.4\% of C atom moves were successful. Figure \ref{fig:dynamics}c includes an inset figure of the unweighted accepted move distributions, highlighting that Cr had a similar contribution to overall acceptance as Ni, despite its lower concentration. This shows that Cr was highly mobile in the mixed metal simulation cell. Additionally, Fe, despite its low atomic concentration (5 at\%), showed notable contributions to metallic moves, indicating its mobility in the mixed lattice.

\subsection{Excess Energy}
\begin{table}[H]
    \centering
    \begin{tabular}{c|ccc|ccc}\hline\hline
       System  & \multicolumn{3}{c|}{+TiB\textsubscript{2}} & \multicolumn{3}{c}{+TiC} \\
          Reference State & B-Mo & B-Ni & B-Ti & C-Cr & C-Ni & C-Ti \\\hline
       $E_{i}$ (eV/atom) & -7.54 & -7.02 & -8.17 & -7.73 & -5.85 & -9.77 \\
       $E_{\text{excess}}$ (meV/atom)  & -29.4 & -41.5 & -15.0 & -38.7 & -82.3 & 8.58 \\
        \hline\hline
    \end{tabular}
    \caption{Excess energy ($E_{\text{excess}}$), as defined in Eq. \ref{eq:excess}, and interstitial energy ($E_i$), as defined in Eq. \ref{eq:mu}, for the doped systems using various reference states to calculate $E_i$.}
    \label{tab:energetics}
\end{table}
To evaluate the thermodynamic stability of B and C within the equilibrated mixed-metal matrix, $E_{\text{excess}}$ was computed relative to three reference states: B-Mo, B-Ti (TiB$_2$), and B-Ni for B, and C-Cr, C-Ti (TiC), and C-Ni for C. Since the $E_{\text{doped}}$ system corresponds to a simulation cell that has undergone MC equilibration, these values reflect the thermodynamic preference for B and C to remain in their SRO'd configurations versus segregating into competing boride and carbide phases. The computed values of $E_{\text{excess}}$ and reference energies $E_i$ for each system are listed in Table~\ref{tab:energetics}.

For B, the lowest $E_{\text{excess}}$ value occurs when B-Ni is the reference state ($-41.5$ meV/atom), indicating that B is most stable in the equilibrated mixed-metal matrix and is unlikely to precipitate to form a compound with Ni. The less negative value for B-Mo ($-29.4$ meV/atom) suggests that B remains stable within the equilibrated matrix, likely due to the formation of the B-Mo-Nb cluster during MC equilibration. The existing coordination environment in the simulation cell provides a favorable energetic state for B, reducing the driving force for segregation into the reference B-Mo phase. However, in a Mo-enriched region, further precipitation into B-Mo remains possible. The least negative $E_{\text{excess}}$ value for TiB$_2$ ($-15.0$ meV/atom) implies that, if sufficient Ti were present, B could segregate into TiB$_2$ rather than remain dissolved in the mixed-metal matrix. Since $E_{\text{excess}}$ is more negative for B-Mo than for TiB$_2$, B is more stable in the mixed-metal SRO state relative to B-Mo than it is relative to TiB$_2$. This suggests that B has a stronger tendency to leave the matrix and form TiB$_2$ than it does to form B-Mo, provided Ti is available. The observed redistribution of B toward Mo and Nb during MC equilibration confirms that Ti content was insufficient to maintain a stable TiB$_2$ motif, leading to the formation of new local ordering with Mo and Nb instead.

For C, the most negative $E_{\text{excess}}$ occurs when C-Ni is the reference state ($-82.3$ meV/atom), meaning that C is highly stable in the mixed-metal matrix (within the large C-Cr motif) and has almost no tendency to precipitate and form a compound with Ni. The less negative excess energy for C-Cr ($-38.7$ meV/atom) suggests that C remains thermodynamically stable within the mixed-metal matrix, likely due to the presence of the large C-Cr motif formed during MC equilibration. Since C has already established favorable interactions with Cr in the simulation cell, the reference C-Cr phase does not provide a significantly lower-energy alternative, reinforcing that C is well accommodated within its existing coordination environment. The only positive excess energy value is for TiC ($+8.6$ meV/atom), indicating that C is destabilized in the Cr-C motif within the mixed-metal matrix relative to TiC and will preferentially segregate into TiC if enough Ti is present. However, despite being initially introduced as TiC, C migrated into the C-Cr cluster during MC equilibration, confirming that Cr coordination is the preferred state in the absence of a stronger Ti enrichment. This observation highlights the role of local chemical composition in regulating carbide network formation, particularly in regions with high Ti content. As indicated by the positive $E_{\text{excess}}$ for TiC, Ti’s strong chemical affinity for C creates a thermodynamic driving force for C to be drawn away from Cr-rich networks to form TiC, potentially altering carbide morphology and leading to a more isolated carbide distribution. A recent experimental study further supports this behavior, which showed that large TiC precipitates can drive the dissolution of Cr\(_{23}\)C\(_6\) carbides, disrupting the carbide network and weakening grain boundary stability \cite{weiMicrostructureEvolutionCreeprupture2023}. As Cr-rich carbides are critical for creep resistance, their destabilization can significantly impact mechanical performance. Additionally, TiC precipitates tend to coarsen along grain boundaries, contributing to stress concentration and reduced creep resistance \cite{weiMicrostructureEvolutionCreeprupture2023}.

The trends in $E_{\text{excess}}$ provide insight into the segregation behavior versus solid solution stability of B and C in the mixed-metal matrix. The most negative $E_{\text{excess}}$ values indicate that B and C are highly stable within the matrix and do not favor segregation into Ni-based borides or carbides. However, the least negative (or positive) $E_{\text{excess}}$ values suggest that in Ti-rich environments, B and C are more likely to segregate out of the mixed-metal matrix and precipitate as TiB$_2$ and TiC. These findings underscore the impact of local chemistry in dictating the stability of interstitial species and reinforce the importance of carefully balancing alloy composition and distribution to control phase formation and enhance mechanical performance.

\subsection{Properties}
\subsubsection{Elastic Properties}
The elastic properties for the B-doped and C-doped systems are given in Fig. \ref{fig:dynamics}d as relative values compared to the undoped system at $T$ = 800 \(^{\circ}\)C. Both doped systems exhibited an improved elastic response to deformation compared to the undoped alloy, with the +TiB\textsubscript{2} system showing slightly greater improvement than the +TiC system. The changes in ductility are characterized by the $C"/E$ ratio, \(\zeta\), and $G/B$ values. The decreased $C"/E$ ratio and \(\zeta\) in both doped systems suggests reduced ductility. A reduced \(\zeta\) indicates that the bonds within the cell were firmer and more prone to stretching rather than bending. This should be anticipated given the formation of new bonds between B-Mo, C-Cr, and C-Nb. Additionally, the increase in the G/B ratio for both doped systems suggests more rigid structures. These findings suggest that, at the microstructrual level, the formation of these clusters promotes a more brittle region in the mixed-metal matrix. While ductility was reduced in the doped systems, according to the Pugh and Pettifor criteria, all systems were still classified as ductile \cite{senkovGeneralizationIntrinsicDuctiletobrittle2021a}.
\begin{figure}[H]
    \centering
    \includegraphics[width=\linewidth]{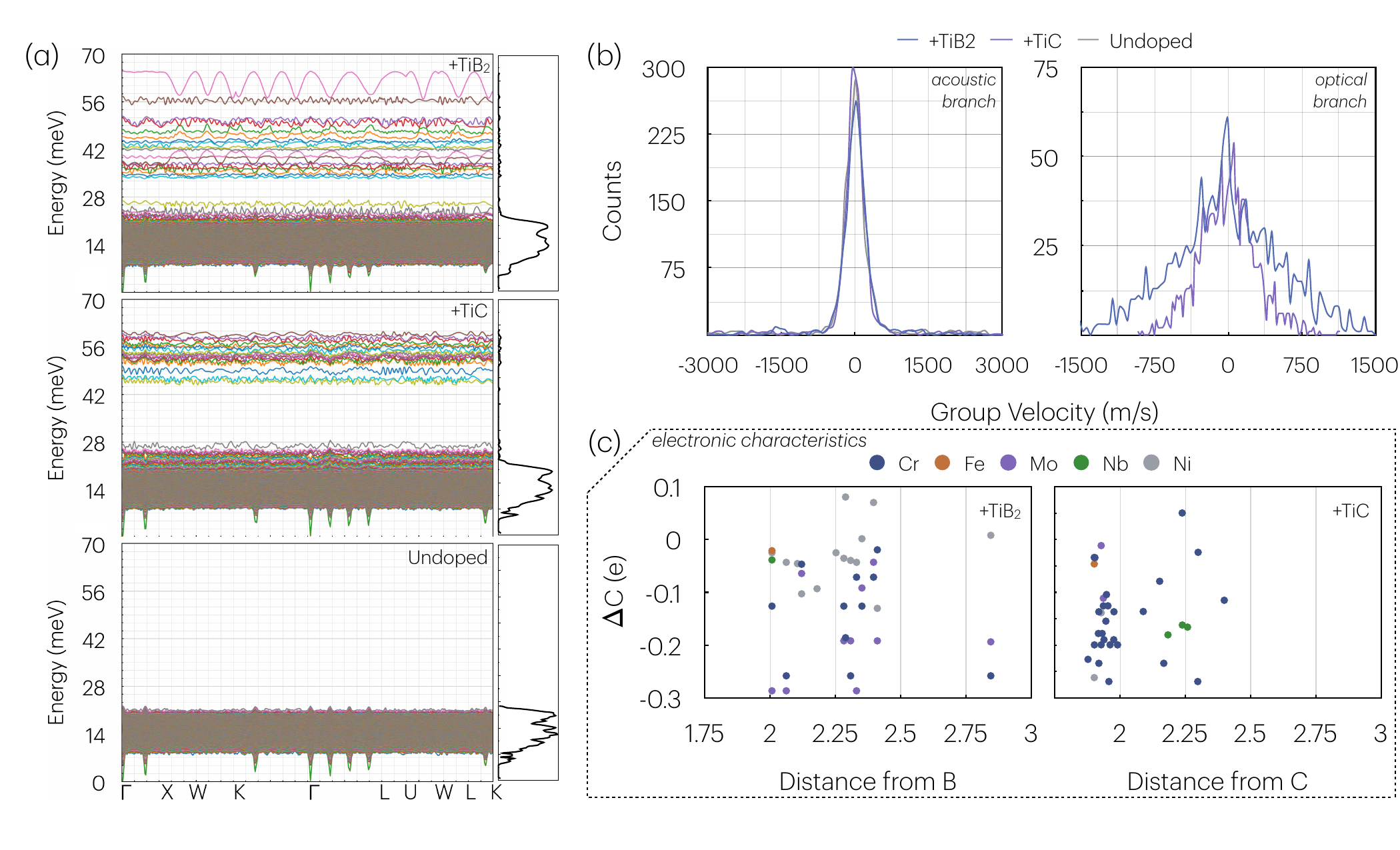}
    \caption{(a) The phonon dispersion curves and (b) acoustic and optical group velocity distribution for the doped and undoped system(s). (c) Difference in charge obtained from the Bader charge analysis. Metal types that were nearest neighbors to a B or C atom are considered here. The values along the $y$-axis represent the change in charge for the neighbor nearest to a B or C atom relative to the average charge of that species obtained from the undoped simulation cell.}
    \label{fig:dispersion-curves}
\end{figure} 

\subsubsection{Phonon Spectrum}

The phonon dispersion curves and group velocity distribution are shown in Fig. \ref{fig:dispersion-curves}a-b. Together, they reveal the vibrational behavior associated with the addition of B or C atoms to the mixed-metal system. A key observation is the development of high-energy optical phonon bands with the addition of B or C. The +TiB\textsubscript{2} structure exhibits a broader range of optical bands and group velocities, attributed to B's lighter atomic mass and its bonding characteristics within the M-B clusters and with neighboring metallic constituents. In contrast, the +TiC system features C atoms bound within a large Cr cluster, resulting in a more narrowly grouped set of optical bands and group velocities. These optical phonon modes reflect how light interstitial dopants modify local bonding environments and vibrational properties.

The +TiC system exhibits a higher density of lower-energy acoustic bands compared to both the undoped and +TiB\textsubscript{2} systems. These acoustic bands correspond to long-wavelength lattice vibrations that govern elastic behavior and stress wave propagation. The formation of M-C clusters perturbs local interatomic coupling, altering the PES and redistributing vibrational energy \cite{stuhlmannSurfacePhononDispersion1989,braunSurfacePhononDispersion1997}. Such perturbations enhance phonon scattering by disrupting atomic regularity \cite{sunSpinphononInteractionsInduced2023, guzmanOpticalAcousticPhonons2023}. The optical phonon modes in the +TiB\textsubscript{2} system suggest a more complex internal lattice dynamics, shaped by varied interactions between B\textsubscript{2} and the surrounding metal atoms. Similar to +TiC, the bonding of B in M-B clusters modifies local vibrational stiffness, redistributing energy across a broader range of modes \cite{stuhlmannSurfacePhononDispersion1989,braunSurfacePhononDispersion1997} which alters phonon transport and energy dissipation \cite{sunSpinphononInteractionsInduced2023, guzmanOpticalAcousticPhonons2023}. The broader optical band range and group velocities in the +TiB\textsubscript{2} system highlight the significant role of B-doping in reshaping vibrational energy distribution and influencing lattice dynamics. The narrowing of group velocity distribution in the +TiC system is a direct consequence of the strong C-Cr bonding within M-C clusters, which confines vibrational energy and limits phonon dispersion. This localization enhances phonon-phonon interactions within the cluster, reducing long-range vibrational coherence. In contrast, the +TiB\textsubscript{2} system shows a more extensive optical band range and a broader distribution of group velocities, indicating a greater degree of vibrational delocalization due to the varied interactions between B\textsubscript{2} and surrounding metallic atoms. 

The phonon group velocity characteristics have macroscopic implications. The narrower velocity spread in the +TiC system suggests lower lattice thermal conductivity due to enhanced phonon scattering within the confined M-C clusters. Conversely, the broader phonon dispersion in +TiB\textsubscript{2} suggests greater thermal dissipation potential, as the extended vibrational modes facilitate energy transport across the lattice. Additionally, the increased phonon scattering in both doped systems may contribute to improved creep resistance, as localized vibrational energy dissipation can hinder dislocation motion. The broader vibrational energy distribution in the +TiB\textsubscript{2} system suggests a more adaptable lattice structure, which may enhance mechanical resilience under extreme conditions. Although phonon scattering at metal-nonmetal pairs may limit direct thermal transport, the redistribution of vibrational energy by both M-B and M-C clusters plays a key role in stabilizing the lattice at high temperatures. These findings suggest that the clustering trends observed in each system influence vibrational stiffness, phonon dynamics, and ultimately, macroscopic material performance.

\subsubsection{Bader Charge Analysis}
Figure \ref{fig:dispersion-curves}c presents the Bader charge analysis for the +TiB\textsubscript{2} and +TiC systems, showing the change in charge (\(\Delta C\)) for various metallic constituents (Cr, Fe, Mo, Nb, Ni) as a function of their distance from B in the +TiB\textsubscript{2} system or C in the +TiC system. The charge changes were calculated relative to the average charge of these elements in the undoped system. In the +TiB\textsubscript{2} system, B\textsubscript{2} formed clusters with neighboring Cr, Mo, and Nb atoms, resulting in charge redistribution among its nearest neighbors. Many Cr and Mo atoms exhibit a decrease in charge within the first nearest neighbor shell. The localized charge accumulation about B\textsubscript{2} could contribute to enhanced mechanical stability and resistance to dislocation motion. The localized bonding about B\textsubscript{2} may also help explain the enhanced high-temperature performance of these alloys, as it reflects strong, directional bonding that is more resistant to thermal and mechanical stresses. In the +TiC system, C atoms were primarily surrounded by Cr atoms, forming a partial M\textsubscript{23}C\textsubscript{6} cluster with a tighter charge distribution, especially to Cr, compared to the TiB\textsubscript{2} system. There is clear evidence of localized electron donation from surrounding Cr atoms to C, leading to the formation of stronger bonds in the M-C cluster. By comparing the density of data points between the 1.75 and 2 $\textrm{\AA}$ regions of the Bader charge plots, it is clear that adding C to the mixed metal matrix had a more profound impact on the electronic environment than adding B atoms did. While the B\textsubscript{2} core accepted charge from its nearest Mo and Cr neighbors, C atoms stacked within the Cr cluster promoted a larger change in the metallic charge density. The affect of charge redistribution and other changes in electronic properties of TM-containing alloys have been linked to altering the mechanical response of both the material's elastic properties and changes in dislocation dynamics \cite{dejongElectronicOriginsAnomalous2015, liImpact$d$bandFilling2017, qiEffectsElectronicStructures2019}. Considering this, both the M-B and M-C structures should be anticipated to impact the dislocation dynamics of the doped alloy relative to its undoped form.
\subsection{Limitations}
Before closing, it is important to highlight the limitations of this computational study. Firstly, the hMCMD simulations were executed with a fixed atomic concentration rather than allowing for variations in atomic fraction through a Grand Canonical MC routine. Secondly, the simulation cell size was relatively limited compared to larger-scale SRO calculations. The impact of system size on WC SRO parameters has been explored in prior work \cite{caoCapturingShortrangeOrder2024}, which demonstrated some uncertainty in metal-metal (M-M') SRO values for a ternary system of comparable size and temperature. This suggests that the observed M-M' SRO tendencies in this study may not be fully converged. However, since this work primarily focuses on the ordering behavior between light interstitials and metal species (i-M SRO) rather than exclusively on M-M’ interactions, there is confidence in the i-M SRO predictions. The clustering trends observed in these simulations align well with experimental EDS imaging of boron-doped Inconel samples and known carbide formations, reinforcing the reliability of the findings. Additionally, these two limitations were made clear in the +TiC simulation where Nb\textsubscript{2}C and Cr\textsubscript{23}C\textsubscript{6} were attempting to form, but were limited in the fact that no new C would be added and there was limited space to segregate within the mixed-metal simulation cell. Larger cells with incoming and outgoing dopant and metal constituents may elucidate new SRO pairs or allow for further development and convergence of the clustering behaviors observed here. While grain boundaries were discussed and some insights were provided to help explain experimental observations, a more thorough examination on how M-B clusters form and behave in true grain boundary conditions should be conducted before any conclusive statements can be made. While the hMCMD simulations emulated ``diffusion'', it did not account for diffusion in the same way that could be done with a kinetic MC (kMC) framework. One benefit to a kMC framework is a physical time associated with each event, related to the activation energy barrier, that can help show how quickly (or slowly) the evolution of the SRO takes. All time-dependent considerations here were in reference to the simulation time rather than the physical time. As it stands, the current methodology did allow for an interesting analysis of undoped and doped Ni-based mixed-metal systems at non-zero temperatures, from atomic ordering and elastic response to phonon and electronic considerations. Nevertheless, future work will focus on refining this routine by addressing these limitations and applying it to new metal matrix composite systems.
\section{Conclusion}
In an effort to bridge the knowledge gap on how interstitial atoms influence the micro- and macroscopic properties of complex alloy systems, this study employed hybrid Monte Carlo Molecular Dynamics (hMCMD) simulations to model undoped and doped Ni-based superalloy cells. Both M-B and M-C clusters formed, aligning with experimental observations and suggesting that this atomic ordering could develop effectively under additive manufacturing conditions. The simulations revealed the formation of small, dynamic Cr(BMo)\textsubscript{2} clusters, a larger Cr\textsubscript{23}C\textsubscript{6} M-C structure, and a Nb\textsubscript{2}C cluster. Crucially, these findings highlighted how interstitial residency impacts the short-range ordering (SRO) of the host alloy, giving rise to new microstructural characteristics, particularly through interactions with transition metals. The addition of B and C improved elastic properties relative to the undoped system. Furthermore, the shared ordering preferences of B and C with Cr, Mo, and Nb suggest potential synergies, as both elements compete for these metals, facilitating the formation of mixed boro-carbides, which have been observed in recent studies.

This study also provided insights into microstructural possibilities that could contextualize recent experimental findings. Specifically, this work demonstrated the aggregation of small, agile B\textsubscript{2} clusters with Cr, Mo, and Nb, which have been experimentally observed at grain boundaries. These results suggest a dual role for B-doping in regulating defect dynamics across different temperature regimes. At low temperatures, B\textsubscript{2} clusters act as diffusion barriers, inhibiting dislocation motion and reducing ductility. Conversely, at high temperatures, these clusters stabilize the microstructure by restricting excessive defect diffusion and dislocation movement along grain boundaries. This behavior may explain the enhanced resistance of In625+TiB\textsubscript{2} to deformation and creep, highlighting the critical role of B in balancing ductility and stability through its localized influence on grain boundaries.

In contrast, C doping promoted the formation of a large M-C cluster, Cr\textsubscript{23}C\textsubscript{6}, accompanied by Nb\textsubscript{2}C. Supported by recent experimental and computational studies, these phases contribute to mechanical reinforcement through bulk matrix stabilization and improved grain boundary properties. Notably, the dissociation of TiC facilitated Cr and Nb clustering, indicating that the addition of C to Ni-based superalloys could enhance the development of stable carbide networks. These networks impede dislocation motion and delay crack propagation. Furthermore, increasing the amount of C to drive further ordering, as observed in this study, may reduce the prevalence of fragmented or poorly distributed carbides, which are known to act as crack initiation sites. These findings emphasize the importance of controlling doping levels and processing conditions to optimize mechanical performance.

The excess energy calculations further reinforce these observations, demonstrating that both B and C exhibit strong thermodynamic stability within their respective short-range ordered configurations. The relatively low $E_{\text{excess}}$ for B-Mo and C-Cr indicates that these interstitials remain energetically favored within the mixed-metal matrix, while the positive $E_{\text{excess}}$ for TiC suggests that, under Ti-rich conditions, segregation into this phases is likely, which may weaken Cr-based carbide networks and promote deleterious TiC precipitates. These results provide additional evidence that interstitial ordering is a key mechanism in stabilizing microstructural features and controlling phase formation in Ni-based superalloys.

This work discussed how the reinforcement mechanisms provided by light interstitial dopants, such as B and C, are intrinsically linked to their interactions with host alloy constituents. These results underscore the importance of light interstitials for tailoring Ni-based superalloys for superior high-temperature performance. More broadly, this work demonstrates that interstitial-induced ordering can serve as a design principle for fine-tuning alloy microstructures, enabling the development of high-performance materials for aerospace and energy generation applications.

\section*{Author Contributions}
\textbf{T.D.D} Writing - original draft, writing – review \& editing, visualization, validation, software, methodology, investigation, formal analysis, data curation. \textbf{E.T.} Writing – review \& editing, validation, experimental data. \textbf{J.S.B.} Writing – review \& editing, validation, experimental data. \textbf{G.D.S.}, \textbf{R.F.} and \textbf{J. Li} Project administration and supervision, writing - review \& editing. All authors contributed to the conceptualization of this project.

\section*{Data Availability}
The hybrid Monte Carlo Molecular Dynamics (hMCMD) routine will be made available at \url{https://github.com/tylerdolezal/hybrid_MCMD}. All data generated from this work and post-processing scripts will be provided at \url{https://github.com/tylerdolezal/comp_mat_sci_paper}.

\section*{Acknowledgments}
J. Li acknowledges support from NSF CMMI-1922206 and DMR-1923976.

\bibliographystyle{ieeetr}

\begin{thebibliography}{10}

\bibitem{fahrenholtzUltrahighTemperatureCeramics2017}
W.~G. Fahrenholtz and G.~E. Hilmas, ``Ultra-high temperature ceramics: {{Materials}} for extreme environments,'' {\em Scripta Materialia}, vol.~129, pp.~94--99, Mar. 2017.

\bibitem{niAdvancesUltrahighTemperature2022}
D.~Ni, Y.~Cheng, J.~Zhang, J.-X. Liu, J.~Zou, B.~Chen, H.~Wu, H.~Li, S.~Dong, J.~Han, X.~Zhang, Q.~Fu, and G.-J. Zhang, ``Advances in ultra-high temperature ceramics, composites, and coatings,'' {\em Journal of Advanced Ceramics}, vol.~11, pp.~1--56, Jan. 2022.

\bibitem{wyattUltrahighTemperatureCeramics2023}
B.~C. Wyatt, S.~K. Nemani, G.~E. Hilmas, E.~J. Opila, and B.~Anasori, ``Ultra-high temperature ceramics for extreme environments,'' {\em Nature Reviews Materials}, pp.~1--17, Dec. 2023.

\bibitem{liuApplicationHighthroughputFirstprinciples2021}
B.~Liu, J.~Zhao, Y.~Liu, J.~Xi, Q.~Li, H.~Xiang, and Y.~Zhou, ``Application of high-throughput first-principles calculations in ceramic innovation,'' {\em Journal of Materials Science \& Technology}, vol.~88, pp.~143--157, Oct. 2021.

\bibitem{farhadizadehMechanicalStructuralThermodynamic2020}
A.~R. Farhadizadeh and H.~Ghomi, ``Mechanical, structural, and thermodynamic properties of {{TaC-ZrC}} ultra-high temperature ceramics using first principle methods,'' {\em Materials Research Express}, vol.~7, p.~036502, Mar. 2020.

\bibitem{wangDesignFe2BbasedDuctile2022}
G.~Wang, X.~Chong, Z.~Li, J.~Feng, and Y.~Jiang, ``Design of {{Fe2B-based}} ductile high temperature ceramics: {{First-principles}} calculations and experimental validation,'' {\em Ceramics International}, vol.~48, pp.~27163--27173, Sept. 2022.

\bibitem{zhangStructuralElectronicMechanical2021}
J.~Zhang, L.~Bao, Z.~Kong, R.~Wang, Y.~Duan, H.~Qi, and M.~Peng, ``Structural, electronic, mechanical, and thermodynamic properties of ultra-high-temperature ceramics {$\alpha$}- and {$\beta$}-{{YAlB4}}: {{A}} first-principles study,'' {\em Ceramics International}, vol.~47, pp.~10079--10088, Apr. 2021.

\bibitem{liuEffectsShortrangeChemical2020}
Y.~Liu, G.-P. Zheng, and M.~Li, ``The effects of short-range chemical and structural ordering related to oxygen interstitials on mechanical properties of {{CrCoFeNi}} high-entropy alloys: {{A}} first-principles study,'' {\em Journal of Alloys and Compounds}, vol.~843, p.~156060, Nov. 2020.

\bibitem{yangIntermediateTemperatureEmbrittlement2024}
H.~Yang, Y.~Gao, Y.~Ding, B.~Zhen, and W.~Wang, ``Intermediate temperature embrittlement and {{Portevin-Le Ch{\^a}telier}} effect of {{Inconel}} 625 alloy caused by carbides,'' {\em Materials Today Communications}, vol.~38, p.~108456, Mar. 2024.

\bibitem{huUltrasonicFabricationProcess2024}
H.~Hu, X.~Wang, J.~Wang, W.~Zhai, and B.~Wei, ``An ultrasonic fabrication process for carbides reinforced {{Inconel}} 718 alloy composites,'' {\em Journal of Alloys and Compounds}, vol.~997, p.~174882, Aug. 2024.

\bibitem{tekogluMetalMatrixComposite2024a}
E.~Teko{\u g}lu, A.~D. O'Brien, J.-S. Bae, K.-H. Lim, J.~Liu, S.~Kavak, Y.~Zhang, S.~Y. Kim, D.~A{\u g}ao{\u g}ullar{\i}, W.~Chen, A.~J. Hart, G.-D. Sim, and J.~Li, ``Metal matrix composite with superior ductility at 800~{$^\circ$}{{C}}: {{3D}} printed {{In718}}+{{ZrB2}} by laser powder bed fusion,'' {\em Composites Part B: Engineering}, vol.~268, p.~111052, Jan. 2024.

\bibitem{desousamalafaiaIsothermalOxidationInconel2020}
A.~M. {de Sousa Malafaia}, R.~B. {de Oliveira}, L.~{Latu-Romain}, Y.~Wouters, and R.~Baldan, ``Isothermal oxidation of {{Inconel}} 625 superalloy at 800 and 1000~{$^\circ$}{{C}}: {{Microstructure}} and oxide layer characterization,'' {\em Materials Characterization}, vol.~161, p.~110160, Mar. 2020.

\bibitem{chenEffectHeatTreatment2020}
L.~Chen, Y.~Sun, L.~Li, and X.~Ren, ``Effect of heat treatment on the microstructure and high temperature oxidation behavior of {{TiC}}/{{Inconel}} 625 nanocomposites fabricated by selective laser melting,'' {\em Corrosion Science}, vol.~169, p.~108606, June 2020.

\bibitem{chenImprovementHighTemperature2020}
L.~Chen, Y.~Sun, L.~Li, and X.~Ren, ``Improvement of high temperature oxidation resistance of additively manufactured {{TiC}}/{{Inconel}} 625 nanocomposites by laser shock peening treatment,'' {\em Additive Manufacturing}, vol.~34, p.~101276, Aug. 2020.

\bibitem{pariziaEffectHeatTreatment2020}
S.~Parizia, G.~Marchese, M.~Rashidi, M.~Lorusso, E.~Hryha, D.~Manfredi, and S.~Biamino, ``Effect of heat treatment on microstructure and oxidation properties of {{Inconel}} 625 processed by {{LPBF}},'' {\em Journal of Alloys and Compounds}, vol.~846, p.~156418, Dec. 2020.

\bibitem{kimHightemperatureTensileHigh2020}
K.-S. Kim, T.-H. Kang, M.~E. Kassner, K.-T. Son, and K.-A. Lee, ``High-temperature tensile and high cycle fatigue properties of inconel 625 alloy manufactured by laser powder bed fusion,'' {\em Additive Manufacturing}, vol.~35, p.~101377, Oct. 2020.

\bibitem{sunHightemperatureOxidationBehavior2020}
Y.~Sun, L.~Chen, L.~Li, and X.~Ren, ``High-temperature oxidation behavior and mechanism of {{Inconel}} 625 super-alloy fabricated by selective laser melting,'' {\em Optics \& Laser Technology}, vol.~132, p.~106509, Dec. 2020.

\bibitem{tripathySurfacePropertyStudy2020}
M.~Tripathy, M.~Munther, K.~Davami, and A.~Beheshti, ``Surface property study of additively manufactured {{Inconel}} 625 at room temperature and 510~{$^\circ$}{{C}},'' {\em Manufacturing Letters}, vol.~26, pp.~69--73, Oct. 2020.

\bibitem{ramenatteComparisonHightemperatureOxidation2020}
N.~Ramenatte, A.~Vernouillet, S.~Mathieu, A.~Vande~Put, M.~Vilasi, and D.~Monceau, ``A comparison of the high-temperature oxidation behaviour of conventional wrought and laser beam melted {{Inconel}} 625,'' {\em Corrosion Science}, vol.~164, p.~108347, Mar. 2020.

\bibitem{huInfluenceHeatTreatments2021}
Y.~Hu, X.~Lin, Y.~Li, S.~Zhang, Q.~Zhang, W.~Chen, W.~Li, and W.~Huang, ``Influence of heat treatments on the microstructure and mechanical properties of {{Inconel}} 625 fabricated by directed energy deposition,'' {\em Materials Science and Engineering: A}, vol.~817, p.~141309, June 2021.

\bibitem{wangEffectMagneticField2021}
Y.~Wang, X.~Chen, Q.~Shen, C.~Su, Y.~Zhang, S.~Jayalakshmi, and R.~A. Singh, ``Effect of magnetic {{Field}} on the microstructure and mechanical properties of inconel 625 superalloy fabricated by wire arc additive manufacturing,'' {\em Journal of Manufacturing Processes}, vol.~64, pp.~10--19, Apr. 2021.

\bibitem{sharifitabarHightemperatureOxidationPerformance2022a}
M.~Sharifitabar, S.~Khorshahian, M.~Shafiee~Afarani, P.~Kumar, and N.~K. Jain, ``High-temperature oxidation performance of {{Inconel}} 625 superalloy fabricated by wire arc additive manufacturing,'' {\em Corrosion Science}, vol.~197, p.~110087, Apr. 2022.

\bibitem{guoEffectB4CContent2024}
C.~Guo, S.~Xu, Z.~Chen, H.~Gao, G.~Jiang, W.~Sun, X.~Wang, and F.~Jiang, ``Effect of {{B4C}} content and particle sizes on the laser cladded {{B4C}}/{{Inconel}} 625 composite coatings: {{Process}}, microstructure and corrosion property,'' {\em Journal of Materials Research and Technology}, vol.~30, pp.~6278--6290, May 2024.

\bibitem{preisEffectLaserPower2024}
J.~Preis, Z.~Wang, J.~Howard, Y.~Lu, N.~Wannenmacher, S.~Shen, B.~K. Paul, and S.~Pasebani, ``Effect of laser power and deposition sequence on microstructure of {{GRCop42}} - {{Inconel}} 625 joints fabricated using laser directed energy deposition,'' {\em Materials \& Design}, vol.~241, p.~112944, May 2024.

\bibitem{luoMicrostructureHightemperatureTribological2024}
J.~Luo, X.~Ma, X.~Huang, Y.~Cao, L.~Pan, X.~Zou, J.~Yi, and L.~Shao, ``Microstructure and high-temperature tribological behaviours of nano-{{HfC}} reinforced {{Inconel}} 625 composite coating by plasma-transferred arc welding,'' {\em Wear}, vol.~554--555, p.~205487, Sept. 2024.

\bibitem{placeholder}
E.~Tekoglu, J.-S. Bae, H.-A. Kim, K.-H. Lim, J.~Liu, T.~D. Dole{\v z}al, S.~Y. Kim, M.~A. Alrizqi, A.~Penn, W.~Chen, A.~J. Hart, J.-H. Kang, C.-S. Oh, J.~Park, F.~Sun, S.~Kim, G.-D. Sim, and J.~Li, ``Superior high-temperature mechanical properties and microstructural features of {{LPBF-printed In625-based}} metal matrix composites,'' {\em Materials Today}, Oct. 2024.

\bibitem{ferraresi2021}
R.~Ferraresi, A.~Avanzini, S.~Cecchel, C.~Petrogalli, and G.~Cornacchia, ``Microstructural, mechanical, and tribological evolution under different heat treatment conditions of inconel 625 alloy fabricated by selective laser melting,'' {\em Advanced Engineering Materials}, vol.~23, p.~2100966, 2021.

\bibitem{song2021}
Y.~Song, X.~L. Jiangkun~Fan, P.~Zhang, and J.~Li, ``Thermal processing map and microstructure evolution of inconel 625 alloy sheet based on plane strain compression deformation,'' {\em Materials}, vol.~14, no.~17, p.~5059, 2021.

\bibitem{wang2023}
F.~Wang, H.~Liu, J.~Li, H.~Wan, L.~Yu, and B.~Liu, ``Microstructure evolution and mechanical properties of inconel 625 foils,'' {\em Journal of Materials Engineering and Performance}, vol.~32, pp.~6576--6587, 2023.

\bibitem{stukowskiVisualizationAnalysisAtomistic2009a}
A.~Stukowski, ``Visualization and analysis of atomistic simulation data with {{OVITO}}--the {{Open Visualization Tool}},'' {\em Modelling and Simulation in Materials Science and Engineering}, vol.~18, p.~015012, Dec. 2009.

\bibitem{metropolisEquationStateCalculations1953}
N.~Metropolis, A.~W. Rosenbluth, M.~N. Rosenbluth, A.~H. Teller, and E.~Teller, ``Equation of {{State Calculations}} by {{Fast Computing Machines}},'' {\em Journal of Chemical Physics}, vol.~21, pp.~1087--1092, June 1953.

\bibitem{thompsonLAMMPSFlexibleSimulation2022a}
A.~P. Thompson, H.~M. Aktulga, R.~Berger, D.~S. Bolintineanu, W.~M. Brown, P.~S. Crozier, P.~J. {in 't Veld}, A.~Kohlmeyer, S.~G. Moore, T.~D. Nguyen, R.~Shan, M.~J. Stevens, J.~Tranchida, C.~Trott, and S.~J. Plimpton, ``{{LAMMPS}} - a flexible simulation tool for particle-based materials modeling at the atomic, meso, and continuum scales,'' {\em Computer Physics Communications}, vol.~271, p.~108171, Feb. 2022.

\bibitem{takamotoUniversalNeuralNetwork2022}
S.~Takamoto, C.~Shinagawa, D.~Motoki, K.~Nakago, W.~Li, I.~Kurata, T.~Watanabe, Y.~Yayama, H.~Iriguchi, Y.~Asano, T.~Onodera, T.~Ishii, T.~Kudo, H.~Ono, R.~Sawada, R.~Ishitani, M.~Ong, T.~Yamaguchi, T.~Kataoka, A.~Hayashi, N.~Charoenphakdee, and T.~Ibuka, ``Towards universal neural network potential for material discovery applicable to arbitrary combination of 45 elements,'' {\em Nature Communications}, vol.~13, p.~2991, May 2022.

\bibitem{Matlantis}
``Matlantis, software as a service style material discovery tool.'' https://matlantis.com/.

\bibitem{MatlantisUserTestimonials}
``Matlantis: {{User Testimonials}}, {{Simulation Case Studies}} \& {{Researches}}.'' https://matlantis.com/cases.

\bibitem{PFPValidationPublic2021}
``{{PFP}} validation for public {{V5}}.0.0.'' https://matlantis.com/news/pfp-validation-for-public-v5-0-0, 2021.

\bibitem{mineComparisonMatlantisVASP2023}
S.~Mine, T.~Toyao, K.-i. Shimizu, and Y.~Hinuma, ``Comparison of {{Matlantis}} and {{VASP}} bulk formation and surface energies in metal hydrides, carbides, nitrides, oxides, and sulfides,'' Apr. 2023.

\bibitem{katoBoronCoordinationThreemembered2024}
T.~Kato, F.~Lodesani, and S.~Urata, ``Boron coordination and three-membered ring formation in sodium borate glasses: A machine-learning molecular dynamics study,'' {\em Journal of the American Ceramic Society}, vol.~107, no.~5, pp.~2888--2900, 2024.

\bibitem{choungRiseMachineLearning2024}
S.~Choung, W.~Park, J.~Moon, and J.~W. Han, ``Rise of machine learning potentials in heterogeneous catalysis: {{Developments}}, applications, and prospects,'' {\em Chemical Engineering Journal}, vol.~494, p.~152757, Aug. 2024.

\bibitem{hisamaMolecularDynamicsCatalystFree2024}
K.~Hisama, K.~V. Bets, N.~Gupta, R.~Yoshikawa, Y.~Zheng, S.~Wang, M.~Liu, R.~Xiang, K.~Otsuka, S.~Chiashi, B.~I. Yakobson, and S.~Maruyama, ``Molecular {{Dynamics}} of {{Catalyst-Free Edge Elongation}} of {{Boron Nitride Nanotubes Coaxially Grown}} on {{Single-Walled Carbon Nanotubes}},'' {\em ACS Nano}, vol.~18, pp.~31586--31595, Nov. 2024.

\bibitem{hinumaNeuralNetworkPotential2024}
Y.~Hinuma, ``Neural {{Network Potential Molecular Dynamics Simulations}} of ({{La}},{{Ce}},{{Pr}},{{Nd}})0.95({{Mg}},{{Zn}},{{Pb}},{{Cd}},{{Ca}},{{Sr}},{{Ba}})0.{{05F2}}.95,'' {\em The Journal of Physical Chemistry B}, vol.~128, pp.~12171--12178, Dec. 2024.

\bibitem{specialmetals_inconel625_2013}
S.~M. Corporation, ``Inconel alloy 625,'' tech. rep., Special Metals Corporation, 2013.
\newblock Technical report on the properties and performance of INCONEL Alloy 625.

\bibitem{renishaw_inconel625_2024}
R.~plc, ``Inconel 625 material data sheet,'' Tech. Rep. H-5800-6792-03-A, Renishaw plc, New Mills, Wotton-under-Edge, Gloucestershire, UK, 2024.
\newblock Technical report on the properties and performance of additively manufactured Inconel 625 using RenAM 500 series.

\bibitem{wangNewInsightsPartitioning2025}
Z.~Wang, C.~Liang, X.~Ding, and D.~Wang, ``New insights into the partitioning behavior of {{Mo}} in nickel-based superalloys and its effect on microstructure using {{CALPHAD-assisted}} phase field modeling,'' {\em Acta Materialia}, vol.~282, p.~120510, Jan. 2025.

\bibitem{mignanelliGammagammaPrimegammaDouble2017}
P.~M. Mignanelli, N.~G. Jones, E.~J. Pickering, O.~M. D.~M. Mess{\'e}, C.~M.~F. Rae, M.~C. Hardy, and H.~J. Stone, ``Gamma-gamma prime-gamma double prime dual-superlattice superalloys,'' {\em Scripta Materialia}, vol.~136, pp.~136--140, July 2017.

\bibitem{goldschmidt4CARBIDES1967}
H.~J. Goldschmidt, ``4 - {{CARBIDES}},'' in {\em Interstitial {{Alloys}}} (H.~J. Goldschmidt, ed.), pp.~88--213, Butterworth-Heinemann, Jan. 1967.

\bibitem{reedSuperalloysFundamentalsApplications2006}
R.~C. Reed, {\em The {{Superalloys}}: {{Fundamentals}} and {{Applications}}}.
\newblock Cambridge: Cambridge University Press, 2006.

\bibitem{zhangSynergyPhaseMC2024}
L.~Zhang, Q.~Yang, J.~Chen, M.~Zhang, and C.~Xiao, ``Synergy of {$\gamma\prime$} phase, {{MC}} carbide and grain boundary phase on creep behavior for nickel-based superalloy {{K439B}},'' {\em Materials Science and Engineering: A}, vol.~915, p.~147261, Nov. 2024.

\bibitem{tekogluStrengtheningAdditivelyManufactured2023}
E.~Teko{\u g}lu, A.~D. O'Brien, J.~Liu, B.~Wang, S.~Kavak, Y.~Zhang, S.~Y. Kim, S.~Wang, D.~A{\u g}ao{\u g}ullar{\i}, W.~Chen, A.~J. Hart, and J.~Li, ``Strengthening additively manufactured {{Inconel}} 718 through in-situ formation of nanocarbides and silicides,'' {\em Additive Manufacturing}, vol.~67, p.~103478, Apr. 2023.

\bibitem{Cowley1950AnAT}
J.~M. Cowley, ``An approximate theory of order in alloys,'' {\em Physical Review}, vol.~77, pp.~669--675, 1950.

\bibitem{cowley1960}
J.~M. Cowley, ``Short- and long-range order parameters in disordered solid solutions,'' {\em Phys. Rev.}, vol.~120, pp.~1648--1657, Dec 1960.

\bibitem{niuExtraelectronInducedCovalent2012}
H.~Niu, X.-Q. Chen, P.~Liu, W.~Xing, X.~Cheng, D.~Li, and Y.~Li, ``Extra-electron induced covalent strengthening and generalization of intrinsic ductile-to-brittle criterion,'' {\em Scientific Reports}, vol.~2, p.~718, Oct. 2012.

\bibitem{senkovGeneralizationIntrinsicDuctiletobrittle2021a}
O.~N. Senkov and D.~B. Miracle, ``Generalization of intrinsic ductile-to-brittle criteria by {{Pugh}} and {{Pettifor}} for materials with a cubic crystal structure,'' {\em Scientific Reports}, vol.~11, p.~4531, Feb. 2021.

\bibitem{naherAbinitioStudyStructural2021a}
M.~I. Naher and S.~H. Naqib, ``An ab-initio study on structural, elastic, electronic, bonding, thermal, and optical properties of topological {{Weyl}} semimetal {{TaX}} ({{X}} = {{P}}, {{As}}),'' {\em Scientific Reports}, vol.~11, no.~1, p.~5592, 2021.

\bibitem{ekumaElasToolV3Efficient2024b}
C.~E. Ekuma and Z.~L. Liu, ``{{ElasTool}} v3.0: {{Efficient}} computational and visualization toolkit for elastic and mechanical properties of materials,'' {\em Computer Physics Communications}, vol.~300, p.~109161, July 2024.

\bibitem{eberhartCauchyPressureGeneralized2012}
M.~E. Eberhart and T.~E. Jones, ``Cauchy pressure and the generalized bonding model for nonmagnetic bcc transition metals,'' {\em Physical Review B}, vol.~86, p.~134106, Oct. 2012.

\bibitem{kleinmanDeformationPotentialsSilicon1962a}
L.~Kleinman, ``Deformation {{Potentials}} in {{Silicon}}. {{I}}. {{Uniaxial Strain}},'' {\em Physical Review}, vol.~128, pp.~2614--2621, Dec. 1962.

\bibitem{kresseEfficientIterativeSchemes1996}
G.~Kresse and J.~Furthm{\"u}ller, ``Efficient iterative schemes for ab initio total-energy calculations using a plane-wave basis set,'' {\em Physical Review B}, vol.~54, pp.~11169--11186, Oct. 1996.

\bibitem{kresseUltrasoftPseudopotentialsProjector1999}
G.~Kresse and D.~Joubert, ``From ultrasoft pseudopotentials to the projector augmented-wave method,'' {\em Physical Review B}, vol.~59, pp.~1758--1775, Jan. 1999.

\bibitem{perdewGeneralizedGradientApproximation1996}
n.~Perdew, n.~Burke, and n.~Ernzerhof, ``Generalized {{Gradient Approximation Made Simple}},'' {\em Physical Review Letters}, vol.~77, pp.~3865--3868, Oct. 1996.

\bibitem{monkhorstSpecialPointsBrillouinzone1976}
H.~J. Monkhorst and J.~D. Pack, ``Special points for {{Brillouin-zone}} integrations,'' {\em Physical Review B}, vol.~13, pp.~5188--5192, June 1976.

\bibitem{henkelman2006fast}
G.~Henkelman, A.~Arnaldsson, and H.~J{\'o}nsson, ``A fast and robust algorithm for bader decomposition of charge density,'' {\em Computational Materials Science}, vol.~36, no.~3, pp.~354--360, 2006.

\bibitem{sanville2007improved}
E.~Sanville, S.~D. Kenny, R.~Smith, and G.~Henkelman, ``An improved grid-based algorithm for bader charge allocation,'' {\em Journal of Computational Chemistry}, vol.~28, no.~5, pp.~899--908, 2007.

\bibitem{tang2009grid}
W.~Tang, E.~Sanville, and G.~Henkelman, ``A grid-based bader analysis algorithm without lattice bias,'' {\em Journal of Physics: Condensed Matter}, vol.~21, no.~8, p.~084204, 2009.

\bibitem{yu2011accurate}
M.~Yu and D.~R. Trinkle, ``Accurate and efficient algorithm for bader charge integration,'' {\em The Journal of Chemical Physics}, vol.~134, no.~6, p.~064111, 2011.

\bibitem{kayhanTransitionMetalBorides2013}
M.~Kayhan, {\em Transition {{Metal Borides}}: {{Synthesis}}, {{Characterization}} and {{Superconducting Properties}}}.
\newblock PhD thesis, Technische Universit{\"a}t Darmstadt, Darmstadt, June 2013.

\bibitem{okadaStructuralInvestigationCr2B31987}
S.~Okada, T.~Atoda, and I.~Higashi, ``Structural investigation of {{Cr2B3}}, {{Cr3B4}}, and {{CrB}} by single-crystal diffractometry,'' {\em Journal of Solid State Chemistry}, vol.~68, pp.~61--67, May 1987.

\bibitem{liuEffectMoAddition2015}
X.~G. Liu, L.~Wang, L.~H. Lou, and J.~Zhang, ``Effect of {{Mo Addition}} on {{Microstructural Characteristics}} in a {{Re-containing Single Crystal Superalloy}},'' {\em Journal of Materials Science \& Technology}, vol.~31, pp.~143--147, Feb. 2015.

\bibitem{wangInfluenceReCr2016}
B.~Wang, J.~Zhang, T.~Huang, H.~Su, Z.~Li, L.~Liu, and H.~Fu, ``Influence of {{W}}, {{Re}}, {{Cr}}, and {{Mo}} on microstructural stability of the third generation {{Ni-based}} single crystal superalloys,'' {\em Journal of Materials Research}, vol.~31, pp.~3381--3389, Nov. 2016.

\bibitem{maPartitioningBehaviorLattice2021}
Z.~Ma, Y.-L. Pei, L.~Luo, L.~Qin, S.-S. Li, and S.-K. Gong, ``Partitioning behavior and lattice misfit of {$\gamma$}/{$\Gamma\prime$} phases in {{Ni-based}} superalloys with different {{Mo}} additions,'' {\em Rare Metals}, vol.~40, pp.~920--927, Apr. 2021.

\bibitem{wangInsightLowCycle2023}
Q.~Wang, Y.~Wu, J.~Chen, J.~Song, C.~Xiao, and X.~Hui, ``Insight into the low cycle fatigue deformation mechanisms of minor element doping single crystal superalloys at elevated temperature,'' {\em Scripta Materialia}, vol.~224, p.~115151, Feb. 2023.

\bibitem{lvStructuralPropertiesPhase2014}
Z.~Q. Lv, F.~Dong, Z.~A. Zhou, G.~F. Jin, S.~H. Sun, and W.~T. Fu, ``Structural properties, phase stability and theoretical hardness of {{Cr23}}-{\emph{x}}{{M}}{\emph{x}}{{C6}} ({{M}} ~ = ~ {{Mo}}, {{W}}; {\emph{x}}~ = ~ 0--3),'' {\em Journal of Alloys and Compounds}, vol.~607, pp.~207--214, Sept. 2014.

\bibitem{yvonKristallstrukturSubcarbideUebergangsmetallen1967}
K.~Yvon, H.~Nowotny, and R.~Kieffer, ``{Die Kristallstruktur der Subcarbide von {\"U}bergangsmetallen},'' {\em Monatshefte f{\"u}r Chemie und verwandte Teile anderer Wissenschaften}, vol.~98, pp.~34--44, Jan. 1967.

\bibitem{liInfluenceCarbidesPores2024}
R.~Li, Y.~Zhang, H.~Niu, H.~Wang, and H.~Wu, ``Influence of carbides and pores on the localized deformation of nickel-based single-crystal superalloys,'' {\em Progress in Natural Science: Materials International}, vol.~34, pp.~562--568, June 2024.

\bibitem{fengEffectPreAddedHfO22024}
H.~Feng, F.~Liu, Q.~Wang, D.~Wang, J.~Song, C.~Xiao, and Y.~Wu, ``Effect of {{Pre-Added HfO2 Inclusions}} on {{Carbide Morphology}} and {{Deformation Behavior}} in {{DZ125 Nickel-Based Superalloy}},'' {\em Metals}, vol.~14, p.~57, Jan. 2024.

\bibitem{suComputationalStructuralModeling2014}
Y.~Su, Z.~Li, L.~Jiang, X.~Gong, G.~Fan, and D.~Zhang, ``Computational structural modeling and mechanical behavior of carbon nanotube reinforced aluminum matrix composites,'' {\em Materials Science and Engineering: A}, vol.~614, pp.~273--283, Sept. 2014.

\bibitem{karamchedHighResolutionElectron2011}
P.~S. Karamched and A.~J. Wilkinson, ``High resolution electron back-scatter diffraction analysis of thermally and mechanically induced strains near carbide inclusions in a superalloy,'' {\em Acta Materialia}, vol.~59, pp.~263--272, Jan. 2011.

\bibitem{singhRolesRefractorySolutes2024}
J.~B. Singh and K.~V. Ravikanth, ``Roles of {{Refractory Solutes}} on the {{Stability}} of {{Carbide}} and {{Boride Phases}} in {{Nickel Superalloys}},'' {\em Journal of Phase Equilibria and Diffusion}, vol.~45, pp.~824--848, Oct. 2024.

\bibitem{theskaReviewMicrostructureMechanical2023}
F.~Theska, W.~F. Tse, B.~Schulz, R.~Buerstmayr, S.~R. Street, M.~{Lison-Pick}, and S.~Primig, ``Review of {{Microstructure}}--{{Mechanical Property Relationships}} in {{Cast}} and {{Wrought Ni-Based Superalloys}} with {{Boron}}, {{Carbon}}, and {{Zirconium Microalloying Additions}},'' {\em Advanced Engineering Materials}, vol.~25, no.~8, p.~2201514, 2023.

\bibitem{kangMicrostructuralAnalysisGrain2024}
B.~Kang, Y.~Lee, J.~Kim, T.~Ha, and Y.~Kim, ``Microstructural {{Analysis}} on {{Grain Boundary}} of {{Boron}}-- and {{Zirconium}}--{{Containing Wrought Nickel-Based Superalloys}},'' {\em Crystals}, vol.~14, p.~290, Mar. 2024.

\bibitem{sheriffQuantifyingChemicalShortrange2024}
K.~Sheriff, Y.~Cao, T.~Smidt, and R.~Freitas, ``Quantifying chemical short-range order in metallic alloys,'' {\em Proceedings of the National Academy of Sciences}, vol.~121, p.~e2322962121, June 2024.

\bibitem{weiOneStepOxides2024}
S.~Wei, Y.~Ma, and D.~Raabe, ``One step from oxides to sustainable bulk alloys,'' {\em Nature}, vol.~633, pp.~816--822, Sept. 2024.

\bibitem{weiMicrostructureEvolutionCreeprupture2023}
L.~Wei, B.~Pan, Y.~Wang, B.~Li, and X.~Jia, ``Microstructure evolution and creep-rupture behaviour of a low-cost {{Fe-Ni-based}} superalloy,'' {\em Materials Technology}, vol.~38, p.~2270865, Dec. 2023.

\bibitem{stuhlmannSurfacePhononDispersion1989}
C.~Stuhlmann and H.~Ibach, ``Surface phonon dispersion in ultrathin nickel films on {{Cu}}(100),'' {\em Surface Science}, vol.~219, pp.~117--127, Sept. 1989.

\bibitem{braunSurfacePhononDispersion1997}
J.~Braun, K.~L. Kostov, G.~Witte, L.~Surnev, J.~G. Skofronick, S.~A. Safron, and {\relax Ch}.~W{\"o}ll, ``Surface phonon dispersion curves for a hexagonally close packed metal surface: {{Ru}}(0001),'' {\em Surface Science}, vol.~372, pp.~132--144, Feb. 1997.

\bibitem{sunSpinphononInteractionsInduced2023}
Q.~Sun, S.~Hou, B.~Wei, Y.~Su, V.~Ortiz, B.~Sun, J.~Y.~Y. Lin, H.~Smith, S.~Danilkin, D.~L. Abernathy, R.~Wilson, and C.~Li, ``Spin-phonon interactions induced anomalous thermal conductivity in nickel ({{II}}) oxide,'' {\em Materials Today Physics}, vol.~35, p.~101094, June 2023.

\bibitem{guzmanOpticalAcousticPhonons2023}
E.~Guzman, F.~Kargar, A.~Patel, S.~Vishwakarma, D.~Wright, R.~B. Wilson, D.~J. Smith, R.~J. Nemanich, and A.~A. Balandin, ``Optical and acoustic phonons in turbostratic and cubic boron nitride thin films on diamond substrates,'' {\em Diamond and Related Materials}, vol.~140, p.~110452, Dec. 2023.

\bibitem{dejongElectronicOriginsAnomalous2015}
M.~{de Jong}, J.~Kacher, M.~H.~F. Sluiter, L.~Qi, D.~L. Olmsted, A.~{van de Walle}, J.~W. Morris, A.~M. Minor, and M.~Asta, ``Electronic {{Origins}} of {{Anomalous Twin Boundary Energies}} in {{Hexagonal Close Packed Transition Metals}},'' {\em Physical Review Letters}, vol.~115, p.~065501, Aug. 2015.

\bibitem{liImpact$d$bandFilling2017}
H.~Li, C.~Draxl, S.~Wurster, R.~Pippan, and L.~Romaner, ``Impact of $d$-band filling on the dislocation properties of bcc transition metals: {{The}} case of tantalum-tungsten alloys investigated by density-functional theory,'' {\em Physical Review B}, vol.~95, p.~094114, Mar. 2017.

\bibitem{qiEffectsElectronicStructures2019}
L.~Qi, ``Effects of electronic structures on mechanical properties of transition metals and alloys,'' {\em Computational Materials Science}, vol.~163, pp.~11--16, June 2019.

\bibitem{caoCapturingShortrangeOrder2024}
Y.~Cao, K.~Sheriff, and R.~Freitas, ``Capturing short-range order in high-entropy alloys with machine learning potentials,'' July 2024.

\end{thebibliography}

\end{document}